\documentclass[aps,prd,
floatfix,nofootinbib,groupedaddress,showpacs,longbibliography]{revtex4-1}

\usepackage{amsmath}
\usepackage{amssymb}
\usepackage{graphicx}

\usepackage{latexsym}
\usepackage{array}
\usepackage{verbatim}

\usepackage{rotating}
\usepackage{color}

\usepackage[utf8]{inputenc}
\usepackage{ulem}

\def\lsim{\mathrel{\raise.3ex\hbox{$<$\kern-.75em\lower1ex\hbox{$\sim$}}}}
\def\gsim{\mathrel{\raise.3ex\hbox{$>$\kern-.75em\lower1ex\hbox{$\sim$}}}}

\newcommand{\be}{\begin{equation}}
\newcommand{\ee}{\end{equation}}

\newlength{\absize}
\def\lsim{\mathrel{\rlap{\raise 2.5pt \hbox{$<$}}\lower 2.5pt
\hbox{$\sim$}}}

\newcommand{\Lumint}{{\cal L}_{\rm int}}

\allowdisplaybreaks

\begin{document}

\title{Updated constraints on $Z'$ and $W'$ bosons decaying
into bosonic and leptonic \\ final states using the Run~2 ATLAS data}

\author{P. Osland}
\email{Per.Osland@uib.no}
\affiliation{Department of Physics and
Technology, University of Bergen, Postboks 7803, N-5020 Bergen,
Norway}
\author{A.~A. Pankov}
\email{pankov@ictp.it}
\affiliation{The Abdus Salam ICTP
Affiliated Centre, Technical University of Gomel, 246746 Gomel,
Belarus}
\affiliation{Institute for Nuclear Problems, Belarusian
 State University, 220030 Minsk, Belarus}
 \affiliation{Joint
 Institute  for Nuclear Research, Dubna 141980 Russia}
\author{I.~A. Serenkova}
\email{Inna.Serenkova@cern.ch}
\affiliation{The Abdus Salam ICTP Affiliated
Centre, Technical University of Gomel, 246746 Gomel, Belarus}

\date{\today}

\begin{abstract}
The full ATLAS Run~2 data set with time-integrated luminosity of 139 fb$^{-1}$ in
the diboson and dilepton channels is used to probe benchmark models with extended gauge sectors: the $E_6$-motivated Grand Unification models, the left-right symmetric $LR$ and the sequential standard model (EGM). These all predict neutral $Z'$ vector bosons, decaying into lepton pairs, $\ell\ell$, or into electroweak gauge boson pairs $WW$, where one $W$ in turn decays semileptonically. 95\% C.L. exclusion limits on the $Z'$ resonance production cross section times branching ratio to electroweak gauge boson pairs and to lepton pairs in
the mass range of $\sim$ 1 -- 6 TeV  are converted to constraints on the $Z$-$Z'$  mixing parameter and the 
heavy resonance mass. We present exclusion regions on the parameter space of the $Z'$ which are significantly extended compared to those obtained from the previous analyses performed with LHC data collected at 7 and 8 TeV in Run~1 as well as at 13 TeV in Run~2 at time-integrated luminosity of 36.1 fb$^{-1}$ and are the most stringent bounds to date.
Also presented, from a similar analysis of electrically charged $W'$ bosons arising in the EGM, which can decay through $W'\to WZ$ and $W'\to \ell\nu$, are limits on the $W$-$W'$ mixing parameter and the charged $W'$ vector boson mass.
\\

\end{abstract}

\maketitle

\section{Introduction} \label{sec:I}
One of the main goals of the physics programme at the Large Hadron Collider (LHC) is to search for new resonant or non-resonant phenomena that become visible in high-energy proton-proton collisions. A prominent possible signature of such phenomena would be the production of a heavy resonance with its subsequent decay into a pair of leptons or into electroweak vector bosons.
Many scenarios beyond the Standard Model (SM) predict such signals. Possible candidates are neutral and charged heavy gauge bosons which are commonly referred to as $Z'$ and $W'$ bosons, respectively \cite{Zyla:2020zbs}.
Strong constraints have already been set on the production of such new heavy particles.

At the LHC, heavy $Z'$ and $W'$
bosons could be observed through their production as $s$-channel
resonances with subsequent leptonic decays
\begin{equation}
pp\to Z^\prime X \to \ell^+\ell^- X, \label{procleptt}
\end{equation}
and
\begin{equation}
pp\to W^\prime X \to \ell\nu X, \label{proclept}
\end{equation}
respectively, where in what follows, $\ell=e,\mu$ unless otherwise stated.
The production of $Z'$ and $W'$ bosons at hadron colliders is expected to be dominated by the Drell-Yan (DY) mechanism, $q\bar{q}/q\bar{q}^\prime\to Z'/W'$.
The Feynman diagrams for the $Z'$ ($W'$) boson production at the parton level and their dilepton and diboson decays are illustrated in  Fig.~\ref{fig1}.
\begin{figure}[!htb]
\refstepcounter{figure} \label{fig1}
 \addtocounter{figure}{-1}
\begin{center}
\includegraphics[height=30mm]{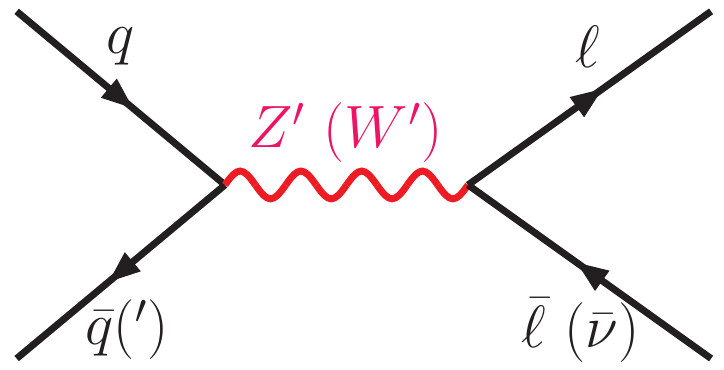} \quad
\includegraphics[height=30mm]{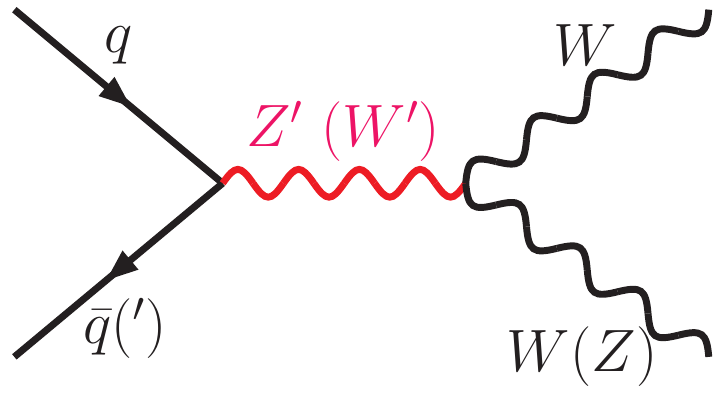}
\end{center}
\vspace*{0.cm} \caption{Parton-level Feynman diagrams for
$Z'$ ($W'$) production with dilepton and diboson decays.
}
\end{figure}

Leptonic final states provide a low-background and efficient experimental signature that  results in  excellent sensitivity to new phenomena at the LHC.
Specifically, these processes (\ref{procleptt}) and (\ref{proclept}) offer the simplest event topology for the discovery of $Z^\prime$ and $W^\prime$ with a large production rate and a clean
experimental signature. These channels offer the most promising discoveries at the LHC \cite{Aad:2019fac, hepdata1725190, CMS:2019tbu, Zucchetta:2019afp,CMS:2018wsn, Aad:2019wvl}.
There have also been many theoretical studies of $Z^\prime$ and $W^\prime$ searches at the high-energy hadron colliders (see, e.g. \cite{Zyla:2020zbs,Altarelli:1989ff,Schmaltz:2010xr,Grojean:2011vu,Jezo:2012rm,Cao:2012ng,Dobrescu:2015yba,
Langacker:2008yv,Erler:2009jh,Hewett:1988xc,Leike:1998wr,Dittmar:2003ir,Osland:2009tn,Godfrey:2013eta,Andreev:2014fwa,Gulov:2018zij,Bandyopadhyay:2018cwu,Bandyopadhyay:2019jzq}).

In the simplest models such as the Sequential Standard Model (SSM) \cite{Altarelli:1989ff} new neutral $Z'_{\rm SSM }$ and charged $W'_{\rm SSM }$ bosons have couplings to fermions that are identical to those of the SM $Z$ and $W$ bosons, but for which the trilinear couplings $Z'WW$ and $W'WZ$ are absent. The SSM has been used as a reference for experimental $Z'$ and $W'$ searches for decades, the results can be re-interpreted in the context of
other models, it is therefore useful for comparing the
sensitivity of different experiments.  Another class of  models considered here are those inspired by Grand Unified Theories (GUT), which are motivated by gauge unification or a restoration of the left--right symmetry violated by the weak interaction. Examples considered in this paper include the $Z^\prime$ bosons of the $E_6$-motivated~\cite{Langacker:2008yv} theories containing
$Z^\prime_\psi$, $Z^\prime_\eta$, $Z^\prime_\chi$; and high-mass neutral bosons of the left-right (LR) symmetric extensions of the SM, based on the $SU(2)_L\bigotimes SU(2)_R\bigotimes U(1)_{B-L}$ gauge group, where $B-L$ refers to the difference between baryon and lepton numbers.

The data we consider were collected with the ATLAS and CMS detectors during the 2015--2018 running period of the LHC, referred to as Run 2 and corresponding to a time-integrated luminosity of 139--140~fb$^{-1}$. The ATLAS  experiment has presented the first search for dilepton resonances  based on the full Run 2 data set \cite{Aad:2019fac,  Aad:2019wvl} and set limits on the $Z'$ and $W'$
production cross sections times branching fraction in the processes (\ref{procleptt}) and (\ref{proclept}), 
$\sigma(pp\to Z'X)\times \text{BR}(Z'\to \ell^+\ell^-)$ and $\sigma(pp\to W'X)\times \text{BR}(W'\to \ell\nu)$, respectively, for $M_{Z'}$ and $M_{W'}$ in the 0.25~TeV -- 6~TeV and 0.15~TeV -- 7~TeV ranges, correspondingly. Recently, similar searches have also been presented by the CMS Collaboration  using  140~fb$^{-1}$ of data recorded at $\sqrt{s}=13$ TeV \cite{CMS:2019tbu}. 
The ATLAS and CMS collaborations set a 95\% confidence level (CL) lower limit on the $Z'$ mass of $\sim$ 4.6 TeV--5.2 TeV depending on the model \cite{Aad:2019fac,CMS:2019tbu}, and 6.0~TeV for the $W'_{\rm SSM}$ \cite{Aad:2019wvl}.

Alternative $Z'$ and $W'$ search channels are the diboson reactions
\begin{equation}
{ pp\to Z^\prime X\to WW X,} \label{procWW}
\end{equation}
and
\begin{equation}
{ pp\to W^\prime X\to WZ X.} \label{procWZ}
\end{equation}

The study of gauge boson pair production offers a powerful test of the
spontaneously broken gauge symmetry of the SM and can be used as a
probe for new phenomena beyond the SM. Specifically, in
contrast to the DY  processes (\ref{procleptt}) and (\ref{proclept})
diboson reactions are not the primary discovery channels, but
can help to understand the origin of new gauge bosons.

As mentioned above, heavy resonances that can decay to gauge boson pairs are
predicted in many scenarios of new physics, including extended
gauge models (EGM) \cite{Altarelli:1989ff,Eichten:1984eu}, models
of warped extra dimensions~\cite{Randall:1999ee,Davoudiasl:2000wi}, technicolour
models~\cite{Lane:2002sm,Eichten:2007sx} associated with
technirho and other technimesons, composite
Higgs models \cite{Agashe:2004rs, Giudice:2007fh}, and the heavy
vector-triplet (HVT) model \cite{Pappadopulo:2014qza}, which generalises a large number of models that predict spin-1 neutral ($Z'$) and charged ($W'$) resonances.  In the SSM, the coupling constants of the
$Z'(W')$ boson with SM fermions are the direct transcription of the
corresponding SM couplings, while the $Z'(W')$ coupling to $WW(WZ)$ is
strongly suppressed, $g_{Z'WW}=0$ and $g_{W'WZ}=0$. This suppression may arise naturally in an EGM: if the new gauge bosons and the SM ones belong to different gauge groups, a vertex such as $Z'WW$ $(W'WZ)$ is forbidden. They
can only be induced after symmetry breaking due to mixing of the gauge eigenstates.
Searches for exotic heavy particles that decay into $WW$ or $WZ$ pairs are complementary to searches in the leptonic channels $\ell^+\ell^-$ and $\ell\nu$ of the processes (\ref{procleptt}) and (\ref{proclept}). Moreover, there are models in which new gauge boson couplings to SM fermions are suppressed, giving rise to a fermiophobic $Z'$ and $W'$ with an enhanced coupling to electroweak gauge bosons~\cite{Zyla:2020zbs, He:2007ge}. It is therefore important to
search for $Z'$ and $W'$ bosons also in the $WW$ and $WZ$ final states.

The properties of possible $Z'$ and $W'$ bosons are also constrained by measurements
of electroweak (EW) processes at low energies, i.e., at energies much below their masses.
Such bounds on the $Z$-$Z'$ ($W$-$W'$) mixing are mostly due to the
constraints on deviation in $Z$ ($W$) properties from the SM predictions.
In particular,  limits from direct hadron production with subsequent diboson decay at the Tevatron \cite{Aaltonen:2010ws} and from  virtual
effects at LEP, through interference or mixing with the $Z$ boson,
imply that any new $Z^{\prime}$ boson is rather heavy and mixes
very little with the $Z$ boson. At LEP and the SLC, the mixing angle is
strongly constrained by very high-precision $Z$ pole experiments
\cite{ALEPH:2005ab}. These include measurements
of the $Z$ line shape, the leptonic branching ratios
as well as leptonic forward-backward asymmetries. The
measurements show that the mixing angles, referred to as  $\xi_{Z\text{-}Z^\prime}$ and $\xi_{W\text{-}W^\prime}$,  between the gauge eigenstates must be smaller than about $10^{-3}$ and $10^{-2}$, respectively \cite{Zyla:2020zbs,Erler:2009jh}.

Previous analyses of the $Z$-$Z'$ and $W$-$W'$ mixing \cite{Osland:2017ema,Bobovnikov:2018fwt,Serenkova:2019zav}
were carried out using the diboson and dilepton production data sets corresponding to the time-integrated luminosity of $\sim$ 36 fb$^{-1}$  collected in 2015 and 2016 with the ATLAS and CMS collaborations at $\sqrt{s}=$ 13~TeV where, in the former case, electroweak $Z$ and $W$ gauge bosons decay into the semileptonic channel \cite{Aaboud:2017fgj} or into the dijet final state \cite{Sirunyan:2017acf}.
The results of the present analysis benefit from the increased size of the data sample, now amounting to an integrated luminosity of  139 fb$^{-1}$
recorded by the ATLAS detector in Run~2 \cite{Aad:2019fbh,hepdata1740685,Aad:2020ddw,hepdata1793572}, almost four times larger than what was available for the previous study.\footnote{In the current analysis, we utilize the full Run~2 ATLAS data set on diboson resonance production \cite{Aad:2019fbh,Aad:2020ddw,hepdata1793572}, rather than that of CMS, as the latter one is unavailable so far.} In addition, further improvement in placing limits on the $Z'$ and $W'$ mass and $Z$-$Z'$ and $W$-$W'$ mixing parameters can be achieved  in semileptonic $WW/WZ$ final states in which one vector boson decays leptonically ($Z\to \ell\ell, \nu\nu$, $W\to \ell\nu$) while the other decays hadronically ($Z/W\to qq$).\footnote{To simplify notation, antiparticles are denoted by the same symbol as the corresponding particles.} Also, here we extend our analysis presented in \cite{Pankov:2019yzr} where we utilized the full Run~2 ATLAS data set for EGM (SSM) to various $Z'$ models, including $E_6$-based $Z_\chi$, $Z_\psi$, $Z_\eta$, and also $Z_{\rm LR}$ boson appearing in models with left-right symmetry. Thus,
our present analysis is complementary to the previous studies \cite{Pankov:2019yzr}.

 We present results as constraints on the relevant $Z$-$Z'$
($W$-$W'$) mixing angle, $\xi_{Z\text{-}Z^\prime}$ ($\xi_{W\text{-}W^\prime}$),  and on the mass $M_{Z'}$ $ (M_{W'}$) and display the
combined allowed parameter space for the benchmark $Z'$ ($W'$) models,
showing also indirect constraints from electroweak precision data.
Previous direct search constraints from the Tevatron and from the LHC with 7 and 8~TeV in Run~1 (where available) are compared to those obtained from the LHC at 13 TeV with the full ATLAS Run~2 data set of time-integrated luminosity of 139 fb$^{-1}$ in the semileptonic \cite{Aad:2020ddw,hepdata1793572} and  fully hadronic ($qqqq$) \cite{Aad:2019fbh} final states.

The paper is structured as follows.
In Sect.~\ref{sect:mixing} we present the theoretical framework, then, in
Sect.~\ref{sect:productionZp} we summarize the relevant cross sections
for the diboson and dilepton production processes (\ref{procWW}) and (\ref{procleptt})  in the narrow-width approximation (NWA). Next, we discuss the relevant $Z'$ widths and branching ratios within the considered benchmark models. Further,  we present an analysis of bounds on  $Z$-$Z'$ mixing from constraints on diboson and dilepton production in the context of the benchmark models with extended gauge sector, employing the most recent searches recorded by the ATLAS (139 fb$^{-1}$) detector  in the semileptonic \cite{Aad:2020ddw,hepdata1793572} and  fully hadronic (referred to as $qqqq$) \cite{Aad:2019fbh,hepdata1740685} final states at the LHC.
Then,  we show the resulting constraints on the $M_{Z'}-\xi_{Z\text{-}Z'}$ parameter space obtained from these processes. Further,  we
collect and compare the indirect constraints obtained from electroweak precision data, direct search constraints derived from the LHC in Run~1 and early Run~2 data.
In Sect.~\ref{sect:productionWp} we present the corresponding analysis of bounds on  $W$-$W'$ mixing, performed in a similar fashion as for the $Z'$, from constraints on diboson and dilepton production processes (\ref{procWZ}) and (\ref{proclept}) in the context of the EGM.
Sect.~\ref{sect:conclusions} presents some concluding remarks.

\section{Mixing and parameters }
\label{sect:mixing}

We consider $Z$--$Z^{\prime}$ mixing within the framework of  models with extended gauge sector such as the $E_6$ models, the LR model and the EGM (see, e.g.
\cite{Altarelli:1989ff,Langacker:2008yv,Erler:2009jh,Hewett:1988xc,Leike:1998wr}).
The mass eigenstates $Z$ and $Z^\prime$ are admixtures of the weak eigenstates $Z^0$ of $SU(2)\times U(1)$ and $Z^{0\prime}$ of the extra $U(1)'$, respectively:
\begin{subequations} \label{Eq:Z12-couplings}
\begin{eqnarray}
&Z& = Z^0\cos\phi + Z^{0\prime}\sin\phi\;, \label{z} \\
&Z^\prime& = -Z^0\sin\phi + Z^{0\prime}\cos\phi\;. \label{zprime}
\end{eqnarray}
\end{subequations}
For each type of $Z'$ boson, defined by its gauge couplings, there are three classes of models, which differ in the assumptions concerning the quantum numbers of the Higgs fields which generate the $Z$-boson mass matrix \cite{Langacker:2008yv,Erler:2009jh}. In each case there is a relation between the $Z^0$-$Z^{0\prime}$ mixing angle $\phi$ and the masses $M_Z$ and $M_{Z'}$ \cite{Langacker:2008yv}:
\begin{equation} \label{phi}
\tan^2\phi={\frac{M_{Z^0}^2-M_Z^2}{M_{Z'}^2-M_{Z^0}^2}}\simeq
\frac{2 M_{Z^0}\Delta M}{M_{Z'}^2}\;,
\end{equation}
where the downward shift $\Delta M=M_{Z^0}-M_Z>0$, and
${M_{Z^0}}$ is the mass of the $Z$ boson in the absence of mixing, i.e., for
$\phi=0$, given by
\begin{equation} \label{MZ0}
{M_{Z^0}}=\frac{M_W}{\sqrt{\rho_0}\cos\theta_W}.
\end{equation}
The mixing angle $\phi$ will play an important role in our analysis. Such mixing effects reflect the underlying gauge symmetry and/or the Higgs sector of the model as the
$\rho_0$ parameter depends on the ratios of Higgs vacuum expectation
values and on the total and third components of weak isospin of the Higgs fields.
We set $\rho_0=1$ here, this corresponds to a Higgs sector with only $SU(2)$ doublets and singlets \cite{Langacker:2008yv}. Once we assume the mass $M_Z$ to be determined experimentally, the mixing depends on two free parameters, which
we identify as $\phi$ and $M_{Z'}$, a parametrization that we will adopt throughout the paper.

This $Z^0$-$Z^{0\prime}$ mixing induces a change in the couplings of the two bosons to fermions. From Eq.~(\ref{Eq:Z12-couplings}), one obtains the vector and
axial-vector couplings of the $Z$ and $Z'$ bosons to fermions:
\begin{subequations} \label{v2}
\begin{alignat}{2}
v_{f} &= v^0_f\cos\phi + v^{0\prime}_f\sin\phi\;, &\quad
a_{f} &= a^0_f \cos\phi + a^{0\prime}_f \sin\phi\;, \label{v1} \\
v^\prime_{f} &= v_f^{0\prime} \cos\phi - v^0_f \sin\phi\;, &\quad
a^\prime_{f} &= a_f^{0\prime} \cos\phi -a^0_f \sin\phi\;,
\end{alignat}
\end{subequations}
with unprimed and primed couplings referring to $Z^{0}$
and $Z^{0\prime}$, respectively, and found, e.g. in~\cite{Leike:1998wr}.

An important property of the models
under consideration is that the gauge eigenstate $Z^{0\prime}$ does
not couple to the $W^+W^-$ pair since it is neutral under
$SU(2)$. Therefore the $W$-pair production is sensitive to a
$Z^\prime$ only in the case of a non-zero $Z^0$--$Z^{0\prime}$ mixing.
From Eq.~(\ref{Eq:Z12-couplings}), one obtains:
\begin{subequations}
\begin{eqnarray}
&& g_{WWZ}=\cos\phi\;g_{WWZ^{0}}\;, \label{WWZ1} \\
&& g_{WWZ'}=-\sin\phi\; g_{WWZ^{0}}\;,\label{WWZ2}
\end{eqnarray}
\end{subequations}
where $g_{WWZ^{0}}=e\cot\theta_W$. Also,
$g_{WW\gamma}=e$.

In many extended models, while the couplings to fermions do
not differ much from those of the SM, the $Z'WW$ coupling is
substantially suppressed with respect to that of the SM. In fact,
in the extended gauge models  the SM trilinear gauge boson
coupling strength, $g_{WWZ^{0}}$, is replaced by $g_{WWZ^{0}} \rightarrow
\xi_{Z\text{-}Z'}\cdot g_{WWZ^{0}}$, where $\xi_{Z\text{-}Z'} \equiv \vert\sin\phi\vert$
(see Eq.~(\ref{WWZ2})) is the mixing factor\footnote{For weak mixing, $\xi_{Z\text{-}Z'}\simeq|\phi|$, and is therefore often referred to as a mixing ``angle''.}.
We will set cross section limits on such $Z'$ as functions of the mass $M_{Z'}$ and $\xi$.

In addition, we study $W$--$W^{\prime}$ mixing in the process  (\ref{procWZ})
within the framework of the EGM model \cite{Altarelli:1989ff,Eichten:1984eu}.
At the tree-level mass mixing may be induced between the electrically charged gauge bosons. The physical (mass) eigenstates of $W$ and $W^\prime$ are admixtures of the weak eigenstates denoted as $\hat{W}$ and $\hat{W'}$, respectively, and obtained  by a rotation of those fields \cite{Zyla:2020zbs,Dobrescu:2015yba}:
\begin{subequations}
\begin{eqnarray}
W^\pm & = & \hat{W}^\pm \cos\theta + \hat{W'}^\pm \sin\theta,  \\
W'^\pm & = & -\hat{W}^\pm \sin\theta + \hat{W'}^\pm \cos\theta,
\end{eqnarray}
\end{subequations}
in analogy with Eq.~(\ref{Eq:Z12-couplings}).
Upon diagonalization  of their mass matrix, the couplings of the observed $W$ boson are shifted from the SM values.

The properties of possible $Z'$ and $W'$ bosons, apart from collider experiments,  are also constrained by measurements of electroweak (EW) processes at low energies, i.e., at energies much below their masses.
Such bounds on the $Z$-$Z'$ ($W$-$W'$) mixing are mostly due to the constraints on the 
deviation in $Z$ ($W$) properties compared to the SM predictions. These
measurements show that the mixing angles $\xi_{Z\text{-}Z^\prime}$ and $\xi_{W\text{-}W^\prime}\,(\equiv \vert\sin\theta\vert)$  between the gauge
eigenstates must be smaller than about $\sim 10^{-3}$ and $10^{-2}$ \cite{Zyla:2020zbs}, respectively.

\section{$Z'$ production and decay in $pp$ collision}
\label{sect:productionZp}
We shall first consider $Z'$ production in some detail, and subsequently turn to the $W'$ case.
In some sense the $Z'$ sector is richer than the $W'$ sector, different models predict different ratios of the vector and axial-vector couplings. The $W'$ models on the other hand, will all be restricted in the choice of pure left-handed couplings to fermions. Among the $Z'$ models, we start out with a detailed discussion of the $\psi$ model.

\subsection{$Z'$ resonant production cross section}

The $Z'$ production and subsequent decay into $WW$ in
proton-proton collisions occurs via  quark-antiquark annihilation  in the $s$-channel.
The  cross section of the process (\ref{procWW}) can at the LHC be
observed through resonant pair production of gauge bosons $WW$.
Using the narrow width approximation (NWA), one can factorize the
process (\ref{procWW}) into the $Z'$ production and its subsequent
decay,
\begin{equation}
\sigma(pp\to Z' X\to WWX)  = \sigma(pp\to Z'X) \times \text{BR}(Z' \to
WW)\;. \label{TotCrWW}
\end{equation}
Here, $\sigma(pp\to Z' X)$ is the  total (theoretical) $Z'$
production cross section and
$\text{BR}(Z' \to WW)=\Gamma_{Z'}^{WW}/\Gamma_{Z'}$ with
$\Gamma_{Z'}$ the total width of the $Z'$. ``Narrow'' refers to the assumption that the natural width of the resonance is smaller than the typical experimental resolution of 5\% of its mass \cite{Aaboud:2016okv,Sirunyan:2017nrt}. This is valid for a large fraction of the parameter space of the considered models.

\subsection{The $Z'$ width}

In the calculation of the total width $\Gamma_{Z'}$ we consider the
following channels: $Z'\to f\bar f$, $W^+W^-$, and $ZH$
\cite{Salvioni:2009mt,Osland:2017ema,Bobovnikov:2018fwt,Pankov:2019yzr}, where $H$ is the SM Higgs boson and $f$ refers to the SM fermions ($f=l,\nu,q$). Throughout the paper we shall
ignore the couplings of the $Z'$ to any beyond-SM particles such as
right-handed neutrinos, which we take to be heavier than $M{_Z'}/2$, as well as to
SUSY partners and any other exotic fermions. 
Such additional states may all together increase the width of the $Z'$
by up to about a factor of five \cite{Kang:2004bz} and hence lower
the branching ratio into a $W^+W^-$ pair by the same factor.

The total width $\Gamma_{Z'}$ of the $Z'$ boson can then be written as
follows:
\begin{equation}\label{gamma2}
\Gamma_{Z'} = \sum_f \Gamma_{Z'}^{ff} + \Gamma_{Z'}^{WW} +
\Gamma_{Z'}^{ZH}.
\end{equation}
The two last terms,  which are often
neglected in studies at low and moderate values of $M_{Z'}$, are due to $Z$-$Z'$
mixing.  For the range of $M_{Z'}$
values below $\sim 3-4$ TeV,  the dependence of
$\Gamma_{Z'}$ on the values of $\xi_{Z\text{-}Z'}$ (within its allowed range) is unimportant.
Therefore, in this mass range, one can approximate the total width as $\Gamma_{Z'} \approx \sum_f
 \Gamma_{Z'}^{ff}$, where the sum runs over SM fermions only.
 The ratios of $\Gamma_{Z'}^{ff}/M_{Z'}$ for the benchmark models
 are summarized in Table~\ref{tab1}. One can appreciate the narrowness
 of the $Z_{Z'}$ pole from this Table~\ref{tab1}.
\begin{table}[htb]
\caption{Ratio $\Gamma_{Z'}^{ff}/M_{Z'}$ for the $\chi, \psi, \eta$,
${\rm LR}$ and EGM  models.}
\begin{center}
\begin{tabular}{|c|c|}
\hline $Z'$  & $\Gamma_{Z'}^{ff}/M_{Z'}$ [\%] \\
\hline $\chi$ & 1.2 \\
\hline $\psi$ & 0.5 \\
\hline $\eta$ & 0.6 \\
\hline ${\rm LR}$ & 2.0 \\
\hline EGM & 3.0 \\
\hline
\end{tabular}
\end{center}
\label{tab1}
\end{table}

However, for larger $Z'$ masses, $M_{Z'}>4$ TeV, there is an
enhancement in the coupling that cancels the suppression due to the tiny $Z$-$Z'$
mixing parameter $\xi_{Z\text{-}Z'}$ \cite{Salvioni:2009mt}. We note that the 
``Equivalence theorem'' \cite{Chanowitz:1985hj} suggests a
value for $\text{BR}(Z'\to ZH)$ comparable to $\text{BR}(Z'\to W^+W^-)$,
 up to electroweak symmetry breaking effects and phase-space factors.  
 Throughout this paper, for definiteness, we adopt a scenario where
both partial widths are comparable, $\Gamma_{Z'}^{ZH}\simeq
\Gamma_{Z'}^{WW}$ for heavy $M_{Z'}$
\cite{Barger:1987xw,Barger:2009xg,Dib:1987ur}.

For all $M_{Z'}$ values of interest for the LHC the width
of the $Z'$ boson is considerably smaller than the experimental
mass resolution $\Delta M$ for which we adopt  the parametrization
in reconstructing the diboson invariant mass of the $W^+W^-$
system, $\Delta M/M\approx 5\% $, as proposed, e.g., in
\cite{Aaboud:2016okv,Sirunyan:2017nrt}.\footnote{This $\Delta M$ should not be confused with that of Eq.~(\ref{phi}).}

The partial width of the $Z'\to W^+W^-$ decay
channel can be written as \cite{Altarelli:1989ff}:
\begin{equation}
\Gamma_{Z'}^{WW}=\frac{\alpha_{\rm em}}{48}\cot^2\theta_W\, M_{Z'}
\left(\frac{M_{Z'}}{M_W}\right)^4\left(1-4\,\frac{M_W^2}{M_{Z'}^2}\right)^{3/2}
\left[ 1+20 \left(\frac{M_W}{M_{Z'}}\right)^2 + 12
\left(\frac{M_W}{M_{Z'}}\right)^4\right]\cdot\xi_{Z\text{-}Z'}^2. \label{GammaWW}
\end{equation}
For a fixed mixing factor $\xi_{Z\text{-}Z'}$ and at large $M_{Z'}$ where
$\Gamma_{Z'}^{WW}$ dominates over $\sum_f \Gamma_{Z'}^{ff}$ the total width increases
rapidly with the mass $M_{Z'}$ because of the quintic
dependence of the $W^+W^-$ mode on the $Z'$ mass as shown in
Eq.~(\ref{GammaWW}). In this case, the $W^+W^-$ mode (together with $Z'\to ZH$) becomes dominant and $\text{BR}(Z' \to W^+W^-)\to 0.5$ (this value arises from the assumption $\Gamma_{Z'}^{ZH}=\Gamma_{Z'}^{WW}$), while the fermionic decay channels ($\Gamma_{Z'}^{ff}\propto M_{Z'}$) are increasingly suppressed. These features are illustrated in Fig.~\ref{br-psi}, where we plot
$\text{BR}(Z'\to W^+W^-)$ and $\text{BR}(Z'\to e^+e^-)$ vs $M_{Z'}$ for the $Z'_\psi$ model.

\begin{figure}[htb]
\refstepcounter{figure} \label{br-psi} \addtocounter{figure}{-1}
\begin{center}
\includegraphics[scale=0.55]{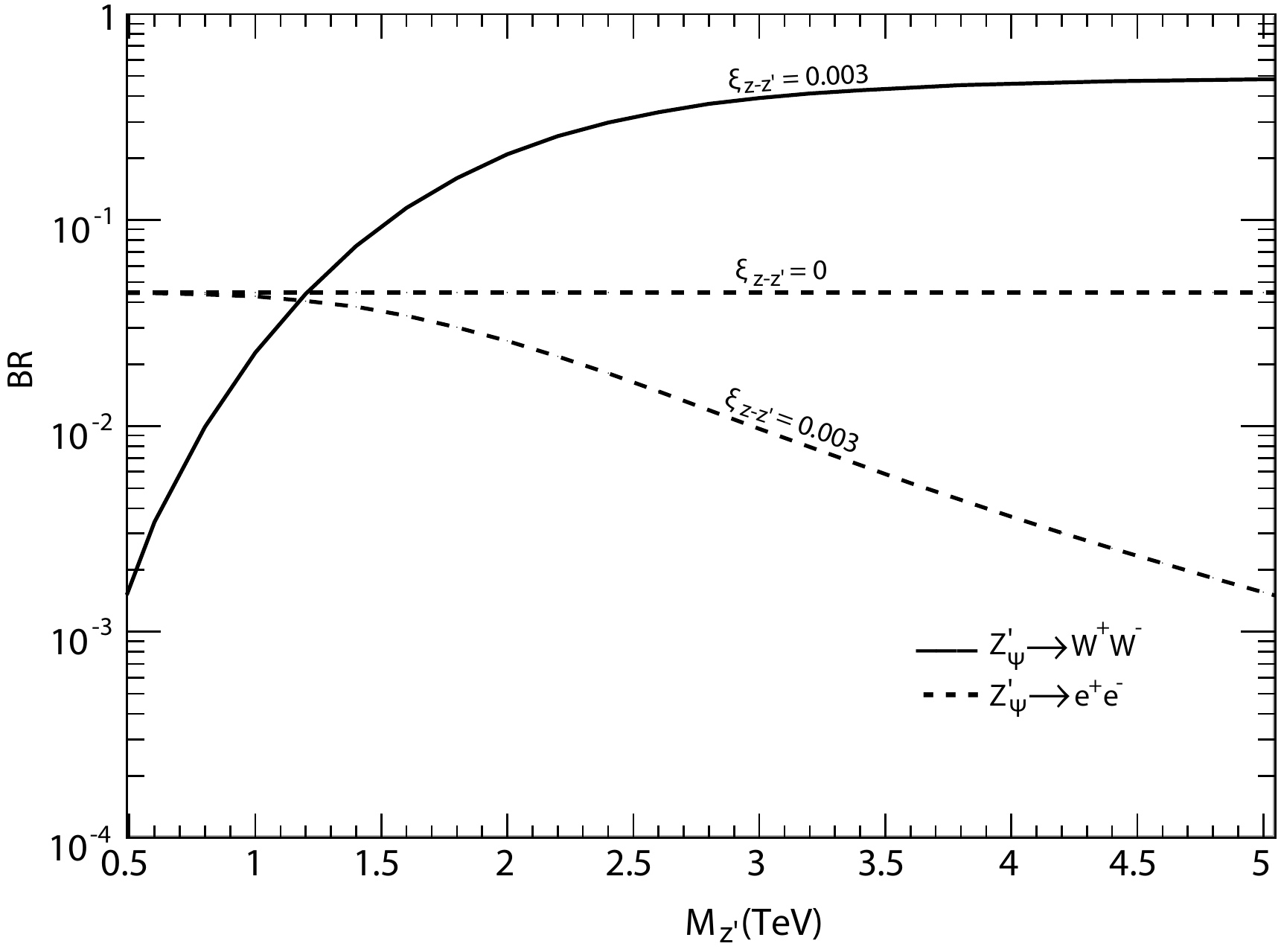}
\end{center}
\caption{
Branching ratio $\text{BR}(Z'\to
W^+W^-)$ (solid) and  $\text{BR}(Z'\to e^+e^-)$ (dashed) vs.\ $M_{Z'}$ in the $Z'_\psi$ model for $Z$-$Z'$ mixing factor $\xi_{Z\text{-}Z'}=0$ and $\xi_{Z\text{-}Z^\prime}=3\cdot 10^{-3}$. It is assumed that $\Gamma_{Z'}^{ZH}=\Gamma_{Z'}^{WW}$.}\label{fig:br-psi}
\end{figure}

\begin{figure}[hbt]
\refstepcounter{figure} \label{sigma-psi} \addtocounter{figure}{-1}
\begin{center}
\includegraphics[scale=0.55]{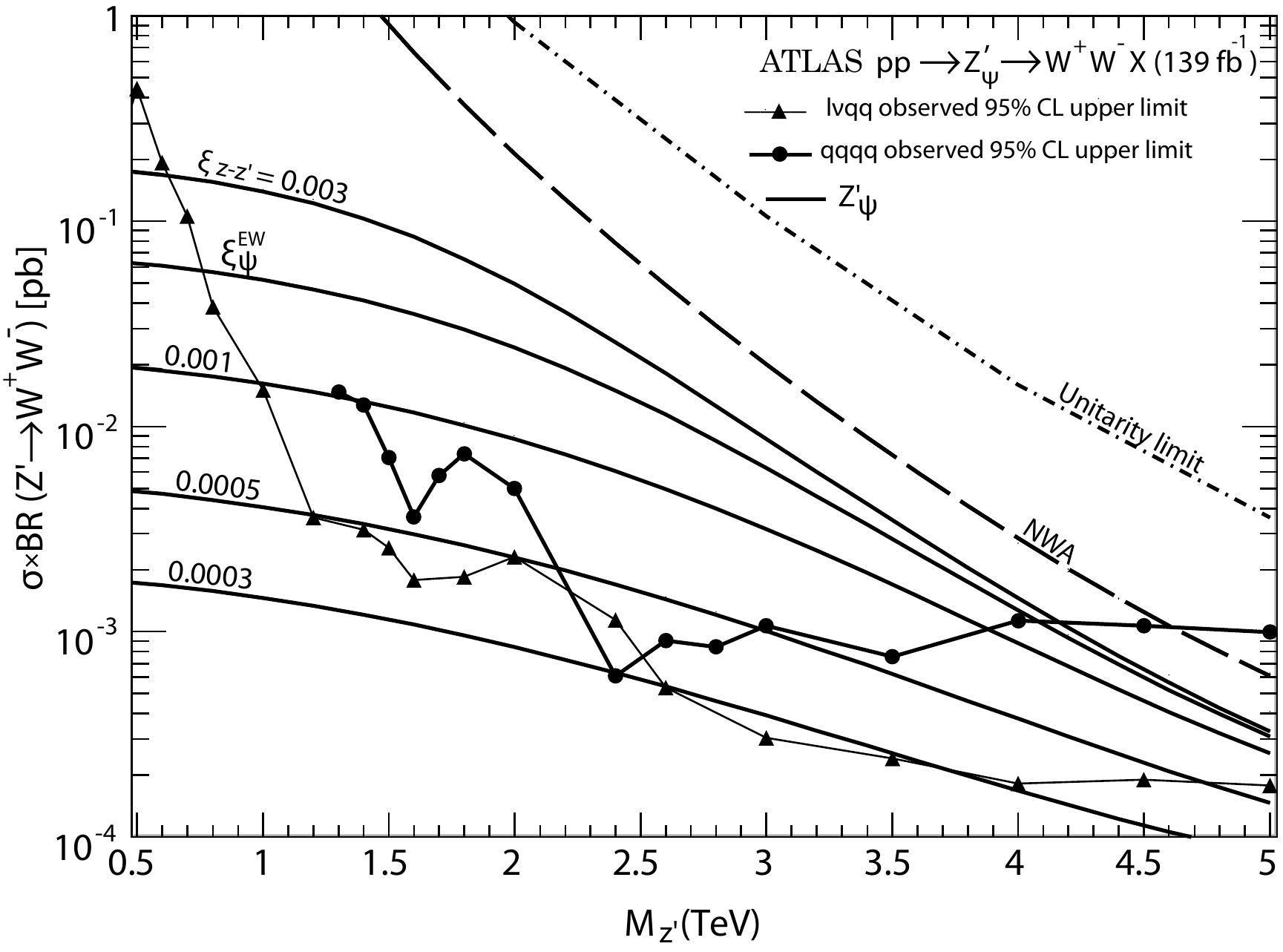}
\end{center}
 \caption{
Observed  $95\%$ C.L. upper limits on the production cross section
times the branching fraction, $\sigma_{95\%}\times \text{BR}(Z'\to
W^+W^-)$, as a function of the $Z'$ mass, $M_{Z'}$, showing ATLAS
data for the semileptonic (thin solid)
\cite{Aad:2020ddw,hepdata1793572} and fully hadronic (thick solid) \cite{Aad:2019fbh} final states for
$139~\text{fb}^{-1}$.
Theoretical production cross sections
$\sigma(pp\to Z'_\psi+X)\times \text{BR}(Z'_\psi\to W^+W^-)$ are shown for mixing factors $\xi_{Z\text{-}Z^\prime}$ ranging from $3\cdot 10^{-3}$ down to
$3\cdot 10^{-4}$.  Also, the cross section solid line  labeled  $\xi^{\rm EW}_{\psi}$
corresponds to the mixing parameter $\xi^{\rm EW}_{Z'}$ indicated in Table~2 for the $Z'_\psi$
model. The area lying below the long-dashed curve labelled NWA  corresponds to the
region where the $Z'$ resonance width is predicted to be less than
5\% of the resonance mass, in which the narrow-width
assumption is satisfied. The lower boundary of the region excluded
by the unitarity constraints is also indicated.
}\label{fig:sigma-psi}
\end{figure}

\subsection{Hadron production and diboson decay of $Z'$}

In Fig.~\ref{sigma-psi}, we consider the full ATLAS Run~2 data set of time integrated luminosity of 139 fb$^{-1}$ and
show the observed  $95\%$ C.L. upper limits on the production cross section times the
branching fraction, $\sigma_{95\%}\times \text{BR}(Z'\to
W^+W^-)$, as a function of the $Z'$ mass, obtained from 
the semileptonic \cite{Aad:2020ddw,hepdata1793572} and  fully hadronic ($qqqq$) \cite{Aad:2019fbh,hepdata1740685} final states.  This allows for a comparison of the sensitivities of the data to mixing parameters and new gauge boson mass. This comparison  demonstrates the dominating sensitivity to $Z'$ of the semileptonic channel with respect to the fully hadronic one, over almost the whole $Z'$ mass range.

Then, for $Z'_\psi$ we compute the LHC production cross section multiplied
by the branching ratio into two $W$ bosons, $\sigma(pp\to Z'_\psi X) \times {\rm
BR}(Z'_\psi\to W^+ W^-)$, as a function of the two parameters
($M_{Z'}$, $\xi_{Z\text{-}Z^\prime}$), and compare it with the limits established by the
ATLAS experiment, $\sigma_{95\%} \times {\rm BR}(Z'\to W^+ W^-)$. The SM backgrounds have been carefully evaluated by the experimental collaborations and accounted for
in $\sigma_{95\%} \times {\rm BR}(Z'\to W^+ W^-)$. Therefore, in our analysis we simulate
only the $Z'_\psi$ signal.

In Fig.~\ref{sigma-psi}, the theoretical production cross section
$\sigma(pp\to Z'_\psi)\times \text{BR}(Z'_\psi\to W^+W^-)$ for $Z'_\psi$
boson is calculated from a dedicated modification of PYHTHIA 8.2
\cite{Sjostrand:2014zea}.
As mentioned above, higher-order QCD corrections to the signal were
estimated using a $K$-factor, for which we adopt a
mass-independent value of 1.9
\cite{Frixione:1993yp,Agarwal:2010sn,Gehrmann:2014fva}. 
These theoretical curves for the cross sections, in descending order,
correspond to values of the $Z$-$Z'$ mixing factor $\xi_{Z\text{-}Z^\prime}$
ranging from $3\cdot 10^{-3}$ and down to $3\cdot 10^{-4}$.
The intersection points of the measured
upper limits on the production cross section with this
theoretical cross section for various values of $\xi_{Z\text{-}Z^\prime}$ give the
corresponding lower bounds on ($M_{Z'}$, $\xi_{Z\text{-}Z^\prime}$), presented
in Fig.~\ref{bounds-psi}.

\begin{figure}[htb!]
\begin{center}
\includegraphics[scale=0.6]{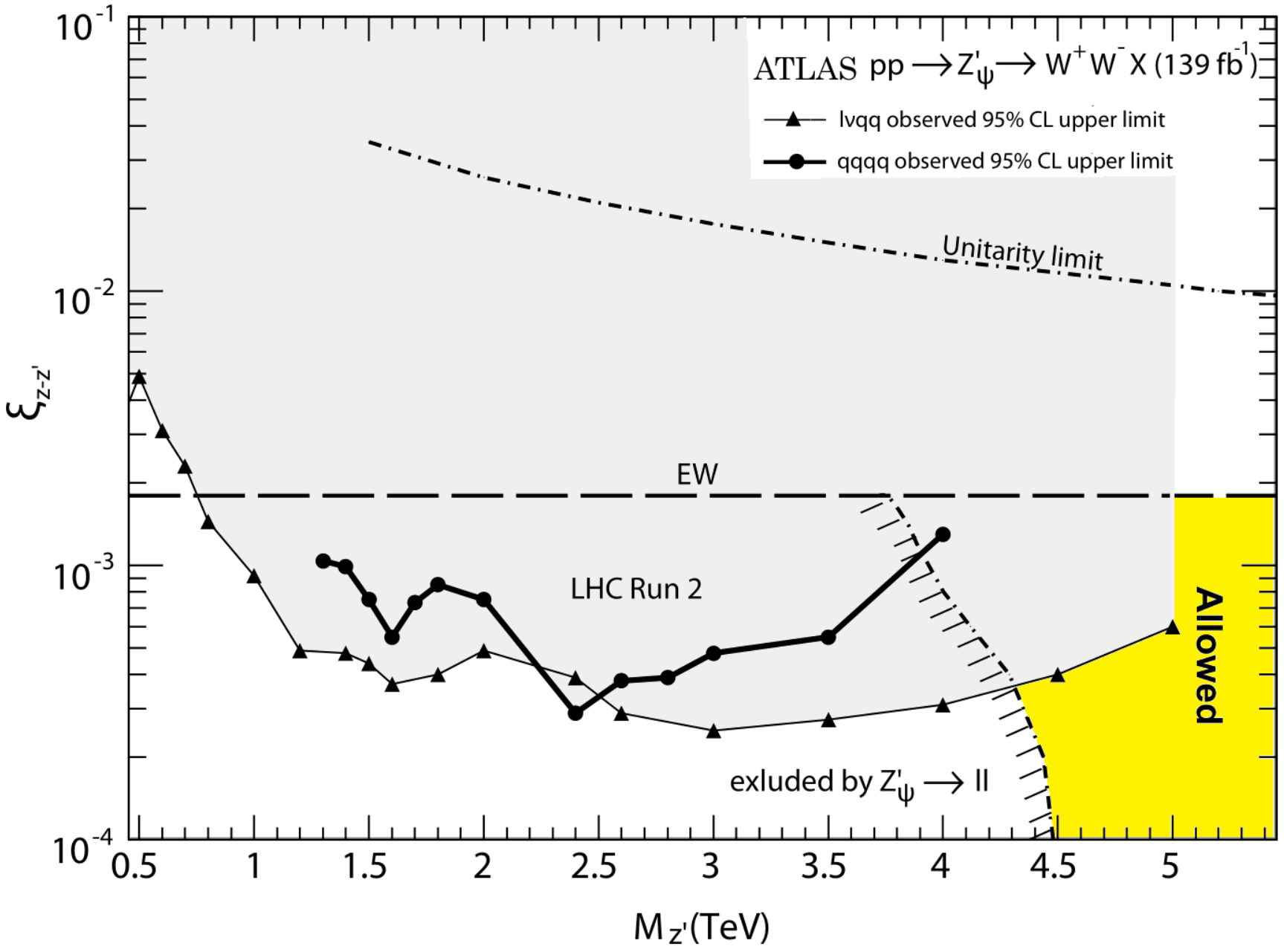}
\end{center}
\caption{The $Z'_\psi$ model: 95\%C.L. exclusion regions in the
two-dimensional ($M_{Z'}$, $\xi_{Z\text{-}Z'}$) plane obtained after
incorporating indirect constraints from
electroweak precision data (horizontal dashed straight line labeled ``EW''), and
direct search constraints from the LHC  search for $pp\to Z'\to WW$ in semileptonic final states using the full Run~2 ATLAS data set. Limits obtained from the hadronic channel $qqqq$  are overlaid  for comparison.
The region above each curve for the $WW$ channel is excluded.
The steep curve labelled ``excluded by $Z'_\psi\to \ell\ell$'' shows the exclusion based on the
dilepton channel $pp\to Z'_\psi\to \ell\ell+X$. The unitarity limit is shown as the dot-dashed curve.
 The overall allowed region is shown as a yellow area.
} \label{fig:bounds-psi}
\label{bounds-psi}
\end{figure}

Different bounds on the $Z'$ parameter space are collected in Fig.~\ref{bounds-psi} for the $Z'_\psi$ model, showing that at high masses, the limits on $\xi_{Z\text{-}Z^\prime}$ obtained from
the full Run~2 data set collected  at $\sqrt{s}=13$ TeV and recorded by the ATLAS detector
are substantially stronger than that derived from the global analysis
of the precision electroweak data \cite{Erler:2009jh}, which is also displayed. 
Limits obtained separately from the individual semileptonic channel $\ell\nu qq$ and the fully hadronic cannel $qqqq$ are shown for comparison. It turns out that the semileptonic channel dominates the sensitivity over almost the whole resonance mass range 0.5~TeV $\leq M_{Z'} \leq$ 5~TeV, while in the rather narrow mass range 2.2~TeV $\leq M_{Z'} \leq$ 2.5~TeV the all-hadronic channel is most sensitive.

\subsection{$Z\text{-}Z'$ mixing effects in dilepton decay of $Z'\to\ell\ell$}
The above analysis was for the diboson process (\ref{procWW}),
employing one of the most recent ATLAS searches for
semileptonic  \cite{Aad:2020ddw,hepdata1793572} and fully hadronic \cite{Aad:2019fbh} final states. Next, we turn to the dilepton production process (\ref{procleptt}), this process gives valuable
complementary information.

We compute the $Z'$ theoretical production cross section at the LHC, $\sigma(pp\to Z'X)$, multiplied by
the branching ratio into two leptons, $\ell\ell$ ($\ell=e,\mu$), i.e.,
$\sigma(pp\to Z'X)\times \text{BR}(Z'\to \ell\ell)$, as a function of $M_{Z'}$,
and compare it with the upper limits established by the
experiment \cite{Aad:2019fac} for $139~\text{fb}^{-1}$.
We make use of the relevant set of tables and figures (including additional results for dielectron and dimuon channels) available at the Durham HepData repository \cite{hepdata1725190}.

Results for $\sigma_{95\%} \times {\rm BR}(Z'\to \ell\ell)$ are shown in
Fig.~\ref{sigmaLL-psi}. To account for
next-to-next-to-leading order (NNLO) effects in the QCD strong
coupling constant, the leading
order (LO) cross sections calculated with PYHTHIA 8.2
\cite{Sjostrand:2014zea} are multiplied by a mass-independent
$K$-factor. The value of the $K$-factor is estimated at a dilepton
invariant mass of $\sim 3.0-4.5$ TeV and found  to be consistent with
unity \cite{Aaboud:2017buh, Sirunyan:2018exx}.

\begin{figure}[hbt]
\refstepcounter{figure} \label{sigmaLL-psi} \addtocounter{figure}{-1}
\begin{center}
\includegraphics[scale=0.6]{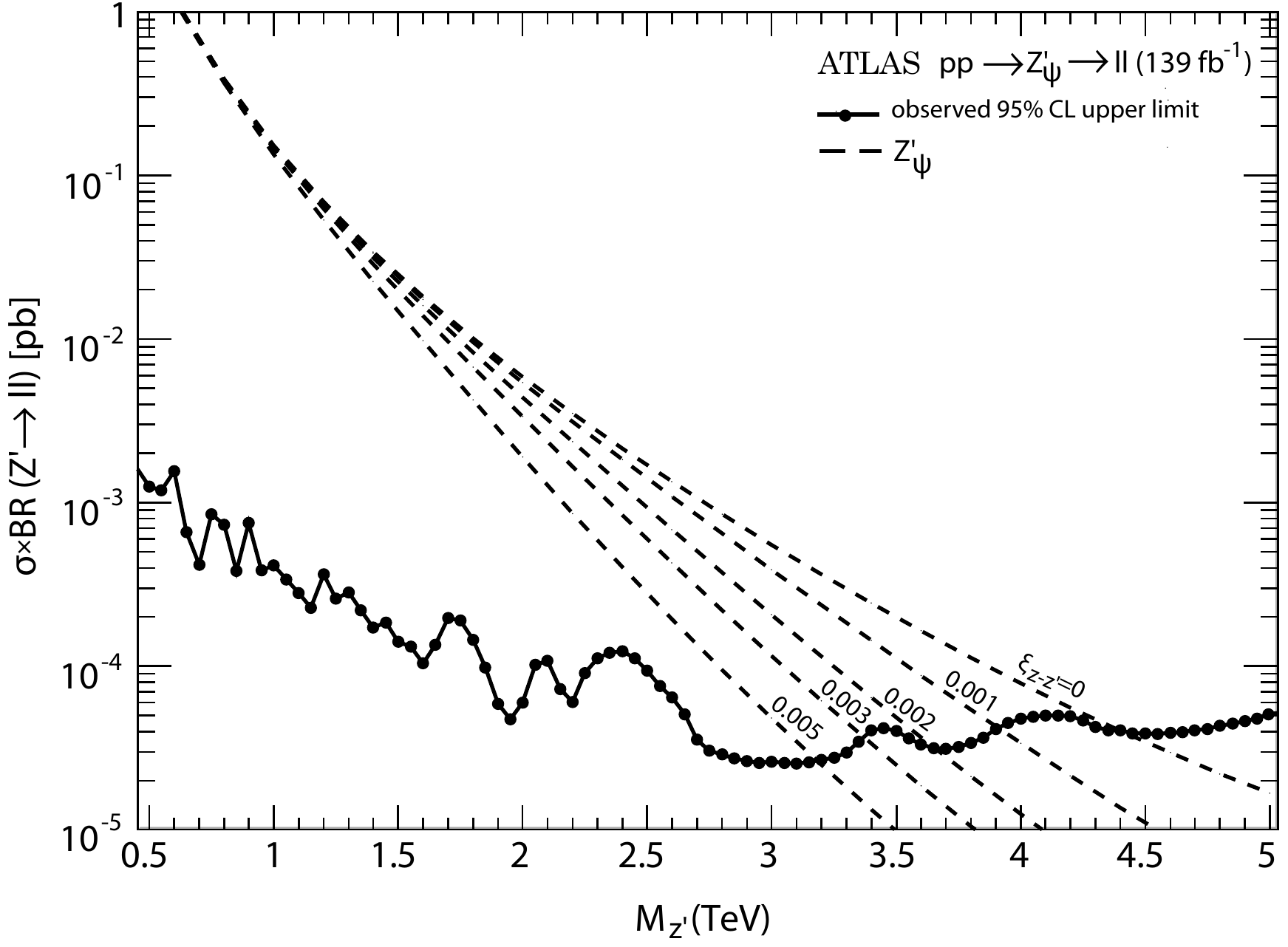}
\end{center}
\caption{
Solid: observed $95\%$
C.L. upper bound on the $Z'$ production cross section times
branching ratio to two leptons, $\sigma_{95\%}\times
\text{BR}(Z'\to \ell\ell)$,  where $\ell=e,\mu$, obtained at the LHC with integrated
luminosity $\Lumint$=139\, fb$^{-1}$ by the ATLAS collaboration
 \cite{Aad:2019fac,hepdata1725190}. Results are shown for the combined dilepton channel.
 Dashed lines: theoretical production cross
section $\sigma(pp\to Z') \times {\rm BR}(Z'\to\ell\ell)$
 for the $Z'$ boson in the $\psi$ model, calculated from PYHTHIA~8.2 with an NNLO $K$ factor.
These curves in descending order correspond to
values of the $Z$-$Z'$ mixing factor $\xi_{Z\text{-}Z'}$ from 0 to 0.005.
} \label{fig:sigmaLL-psi}
\end{figure}

For illustrative purposes we show theoretical production cross sections
$\sigma(pp\to Z'X)\times \text{BR}(Z'\to \ell\ell)$ for the $\psi$ model $Z'$, given by the
dashed curves in Fig.~\ref{sigmaLL-psi}. These curves, in descending order, correspond to values of the mixing factor $\xi_{Z\text{-}Z'}$ from 0 to $5\cdot 10^{-3}$.
Qualitatively, the decrease of the theoretical cross section with
increasing values of $\xi_{Z\text{-}Z'}$ can be understood as follows: For
increasing $\xi_{Z\text{-}Z'}$, the $Z'\to W^+W^-$ mode will at high mass $M_{Z'}$
become more dominant (as illustrated in Fig.~\ref{br-psi}), and $\text{BR}(Z'\to \ell\ell)$
will decrease correspondingly.
Notice also, that applying a mass-dependent $K$-factor
(which for this process is less than 1.04), the $\psi$
model mass limit of the $Z'$ changes by only $\sim{\cal O}$(50 GeV),
justifying the use of the simpler mass-independent $K$-factor
\cite{Aaboud:2017buh, Sirunyan:2018exx}.

\begin{table}[htb]
\caption{Observed 95\% C.L. lower mass limits on $M_{Z'}$ for different
$Z^{\prime}$ gauge models from $pp\to Z^\prime \to \ell\ell X$ taking into account the effect of potential $Z\text{-}Z'$ mixing.}
\begin{center}
\begin{tabular}{|c|c|c|c|}
\hline
Model & mixing parameter & $M_{Z'}$ (TeV) lower limits  \\ \hline
 $Z_\psi^{\prime}$ & no mixing & 4.5  \\
 & $\xi_{Z\text{-}Z'}^{\rm EW}=1.8\cdot 10^{-3}$  & 3.8   \\ \hline\hline
  $Z_\eta^{\prime}$ & no mixing & 4.6  \\
 & $\xi_{Z\text{-}Z'}^{\rm EW}=4.7\cdot 10^{-3}$  & 3.3   \\ \hline\hline
$Z_\chi^{\prime}$ & no mixing  & 4.8  \\
 & $\xi_{Z\text{-}Z'}^{\rm EW}=1.6\cdot 10^{-3}$  & 4.2   \\ \hline\hline
$Z_{\rm LR}^{\prime}$ &  no mixing  & 4.9  \\
 & $\xi_{Z\text{-}Z'}^{\rm EW}=1.3\cdot 10^{-3}$  & 4.5   \\ \hline\hline
$Z_{\rm EGM}^{\prime}$ & no mixing  & 5.1  \\
 & $\xi_{Z\text{-}Z'}^{\rm EW}=2.6\cdot 10^{-3}$  & 4.4   \\ \hline
\end{tabular}
\end{center}
\label{Tab:disc}
\end{table}

\begin{figure}[t]
\begin{center}
\includegraphics[scale=0.42]{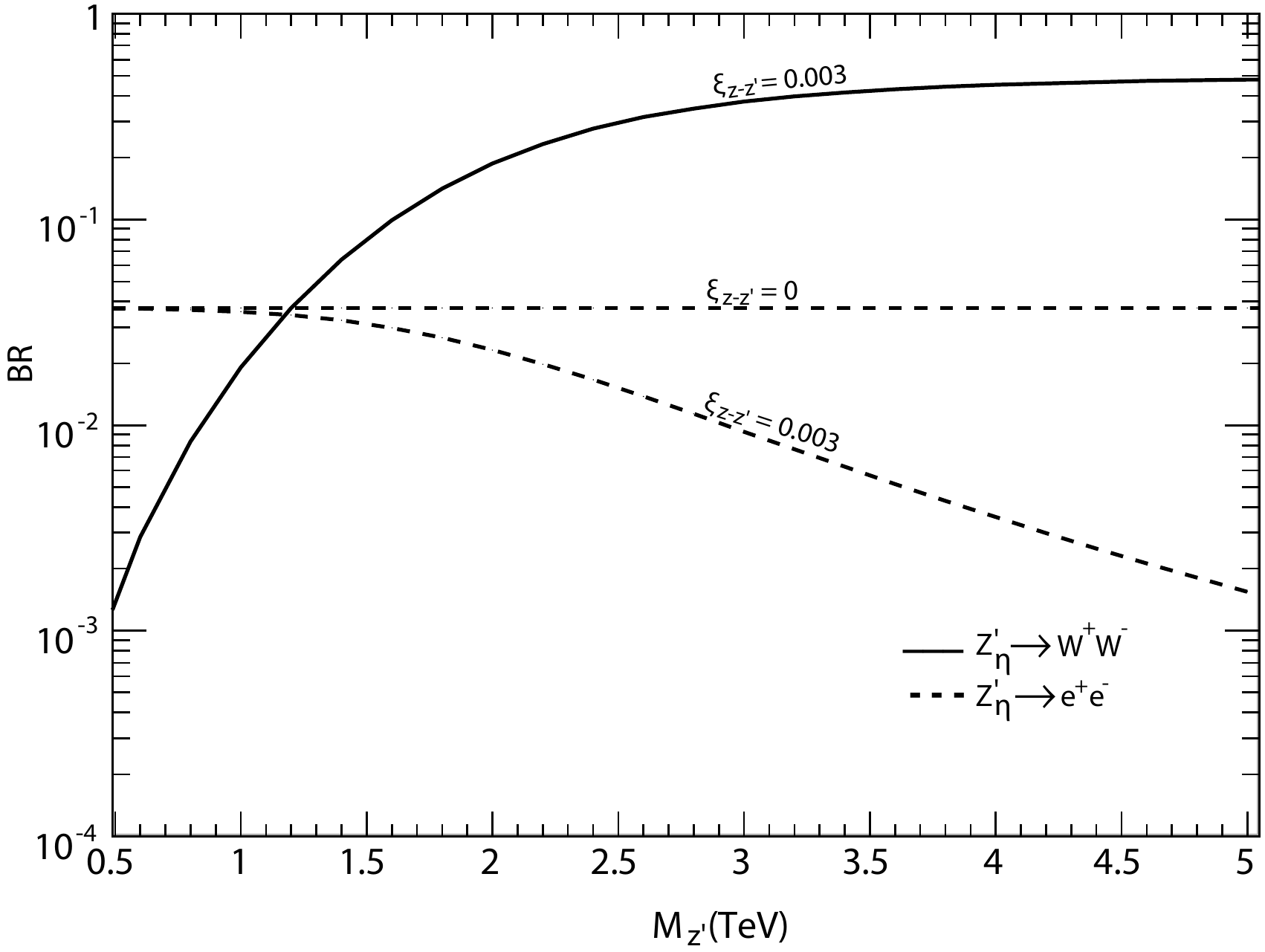}
\includegraphics[scale=0.42]{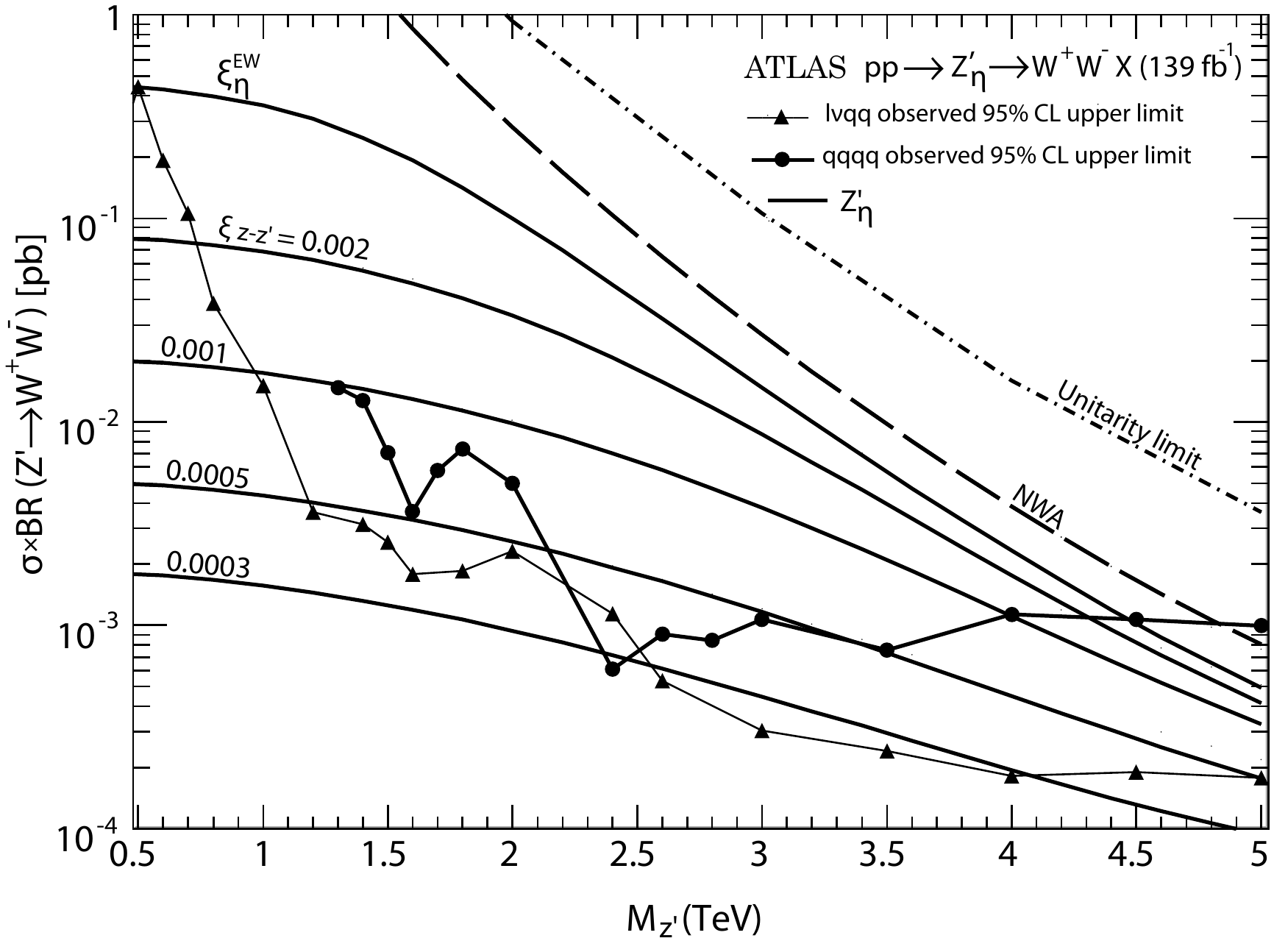}
\includegraphics[scale=0.42]{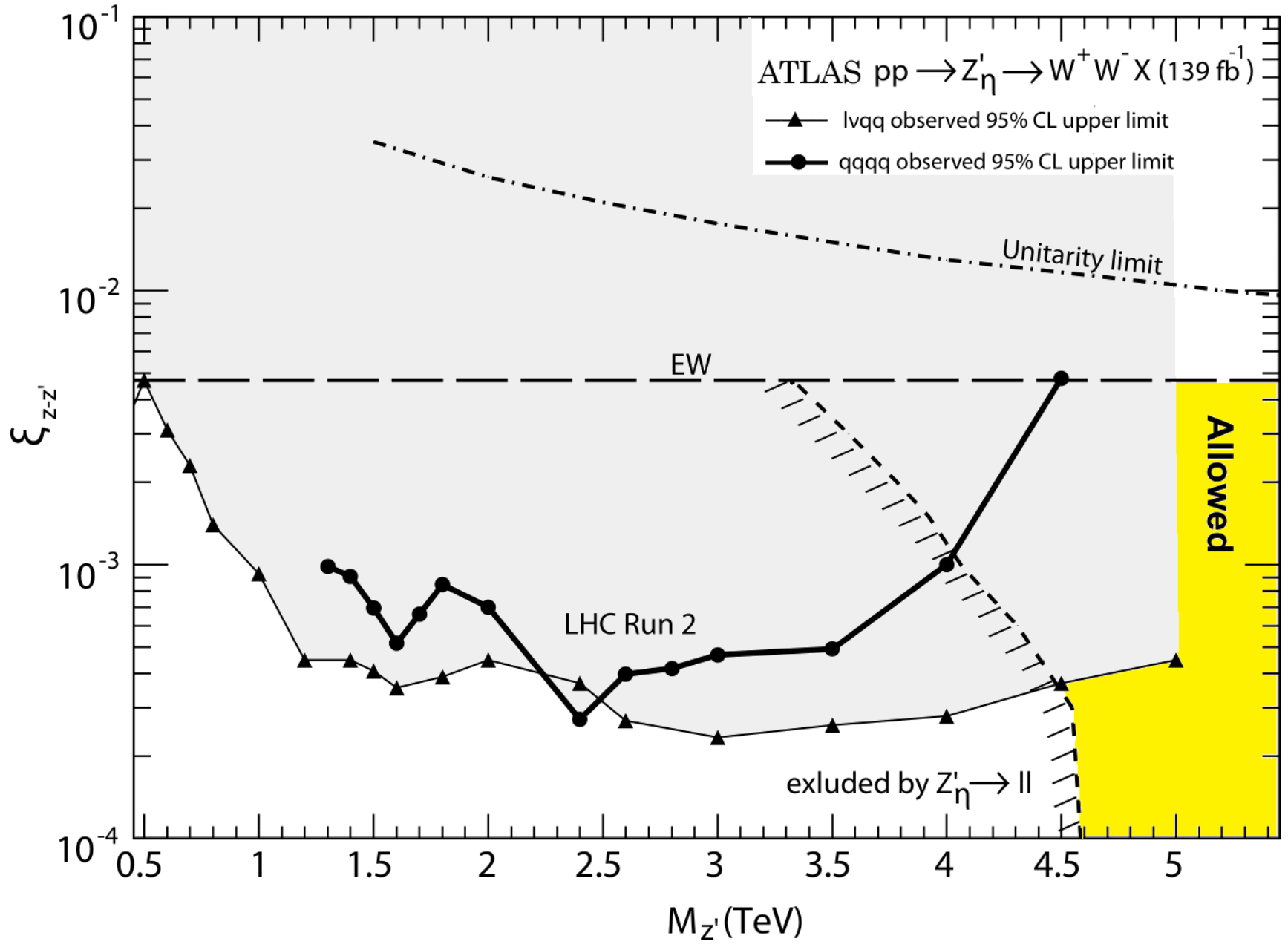}
\includegraphics[scale=0.42]{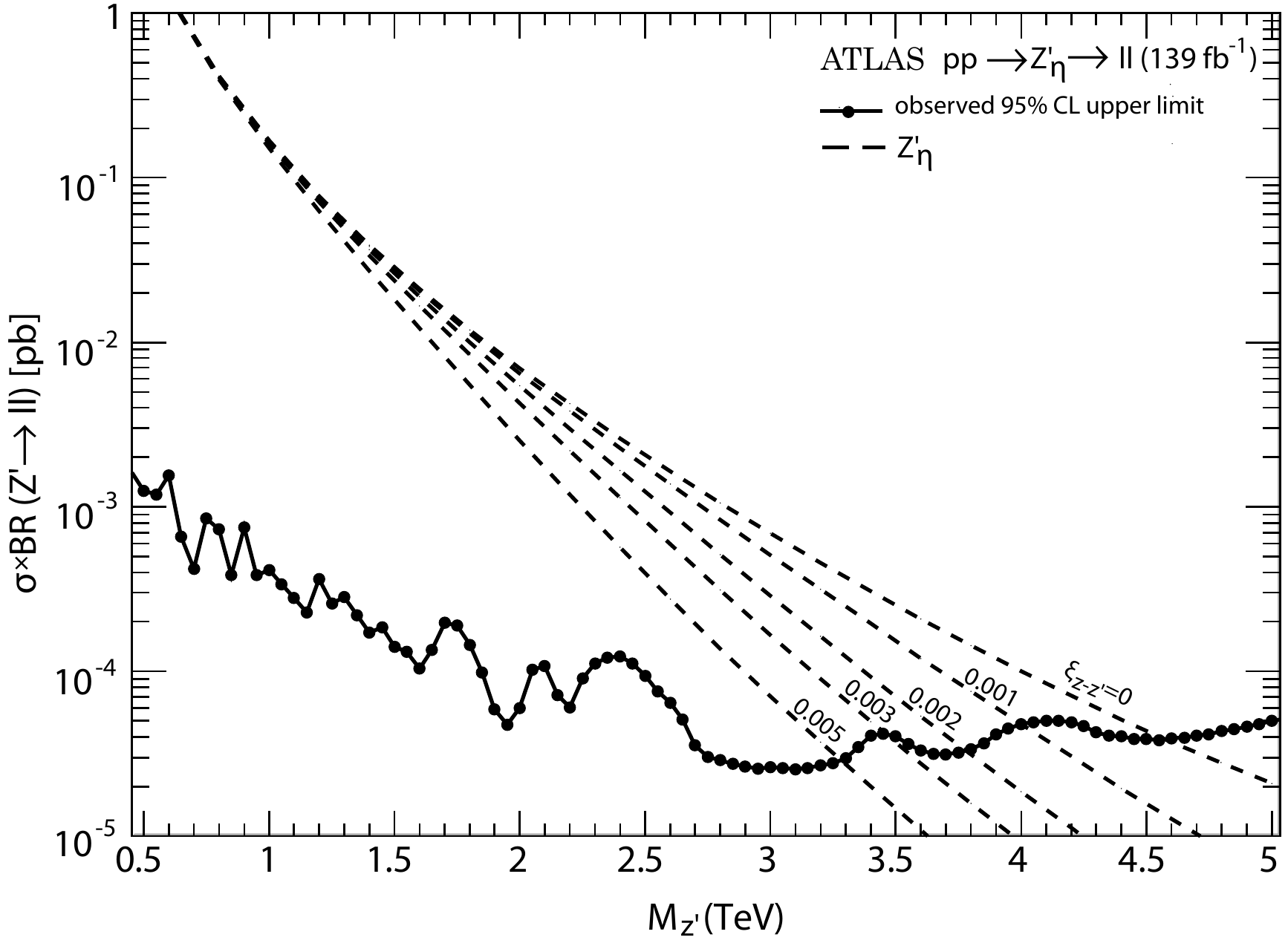}
\end{center}
\caption{The $Z'_\eta$ model: top-left, top-right, bottom-left, bottom-right panels: analogous to Figs.~\ref{fig:br-psi}, \ref{fig:sigma-psi}, \ref{fig:bounds-psi}, \ref{fig:sigmaLL-psi}, respectively.
}
\label{Fig:eta}
\end{figure}

\begin{figure}[t]
\begin{center}
\includegraphics[scale=0.42]{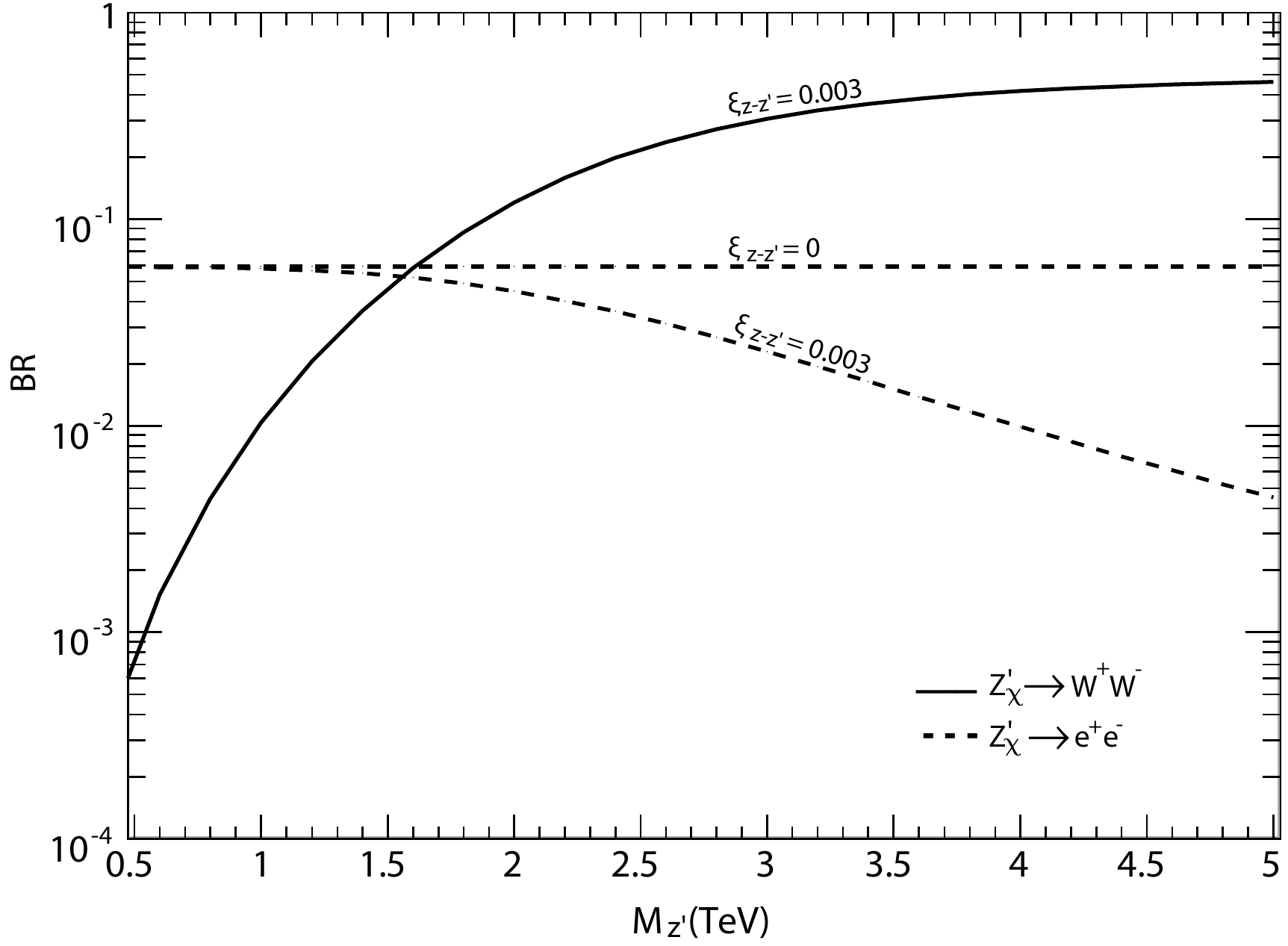}
\includegraphics[scale=0.42]{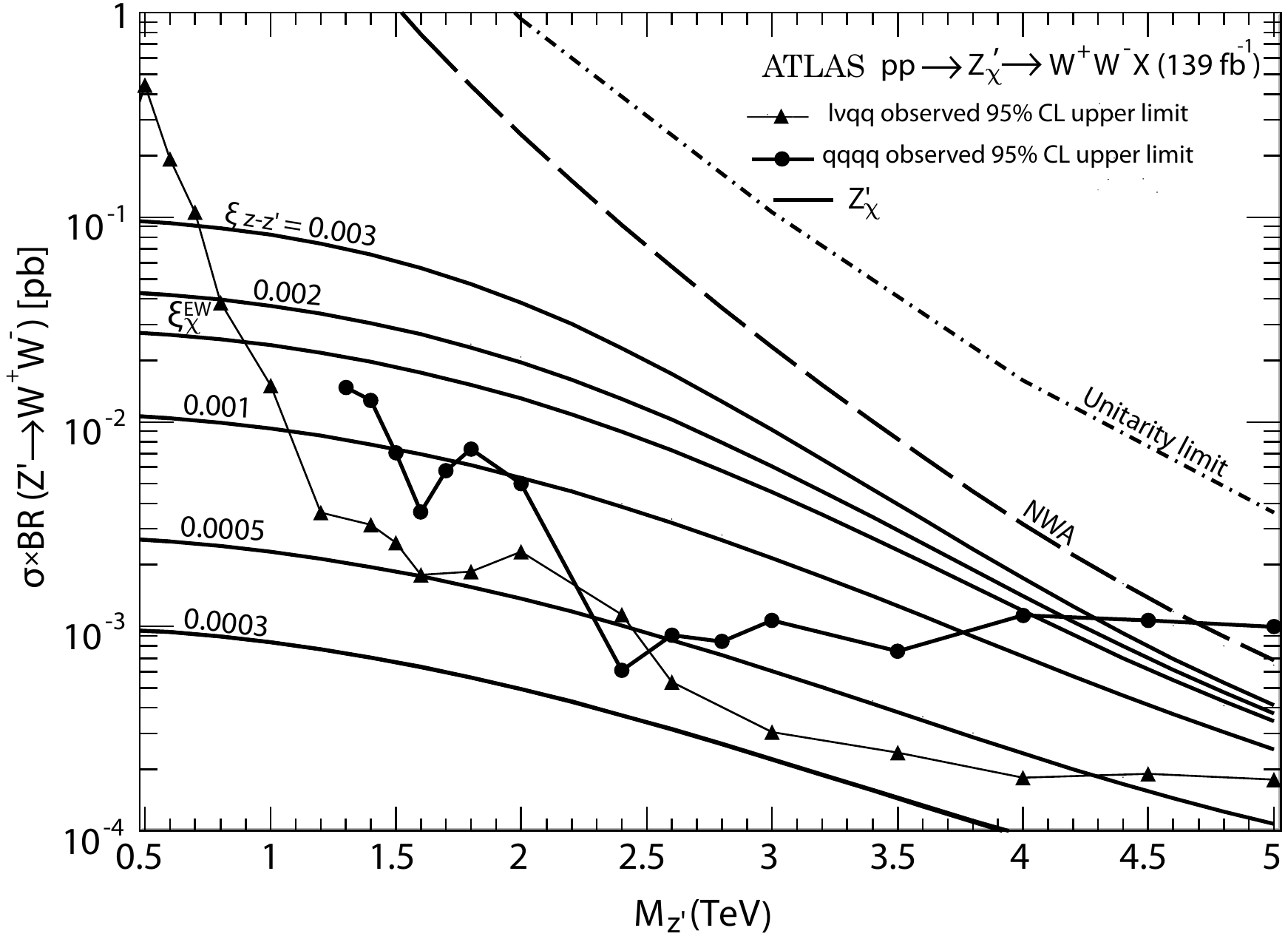}
\includegraphics[scale=0.42]{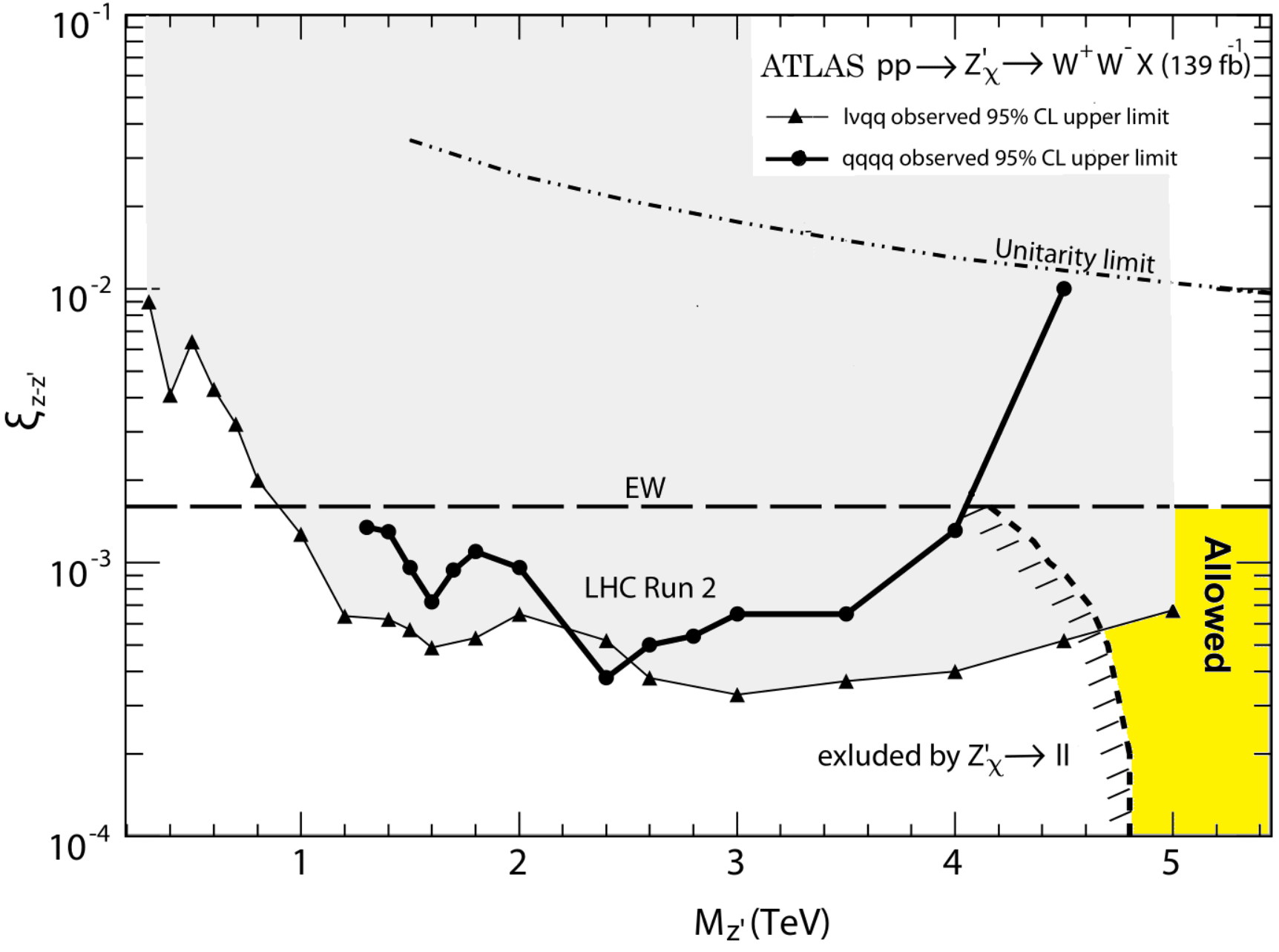}
\includegraphics[scale=0.42]{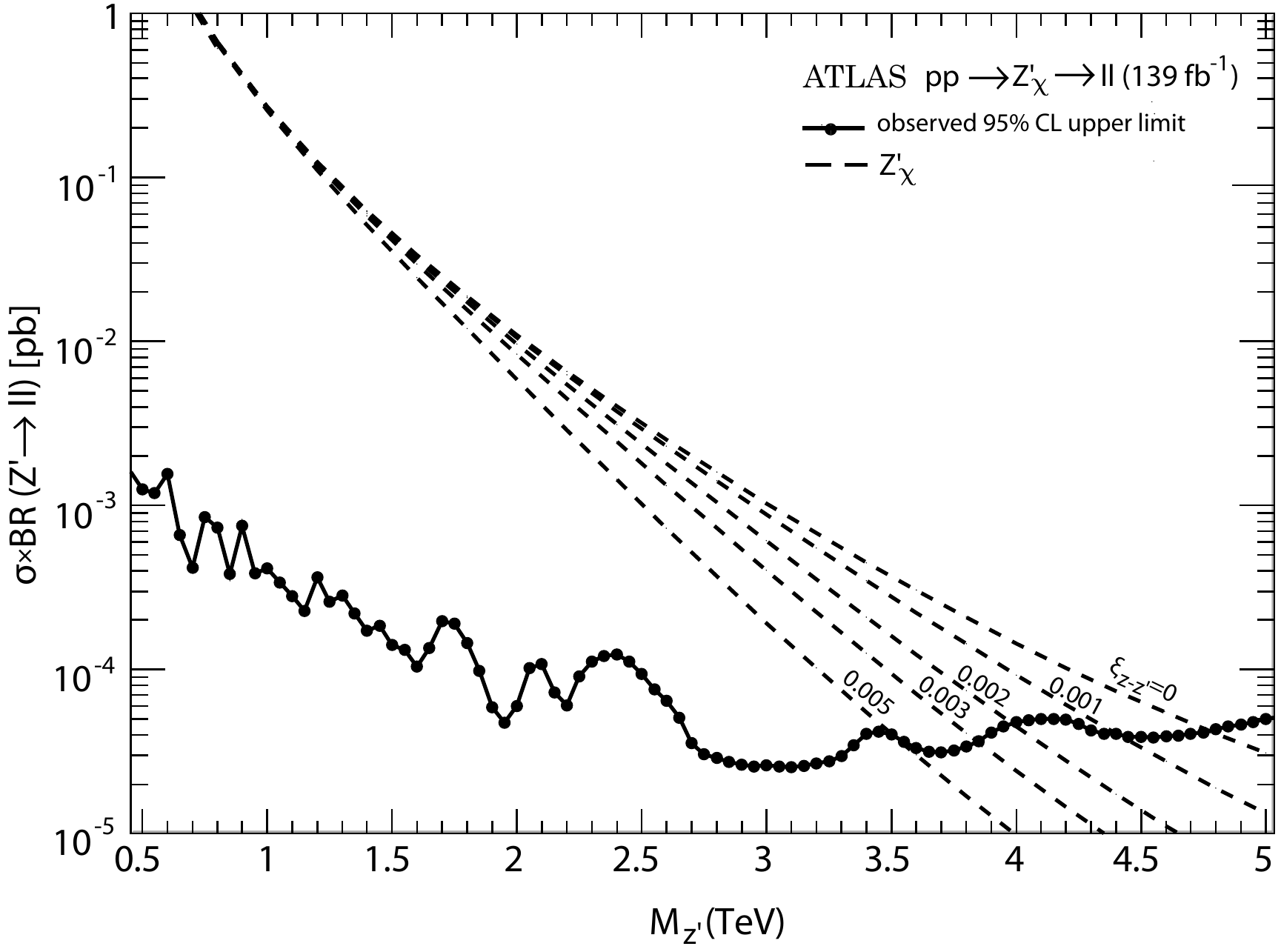}
\end{center}
\caption{The $Z'_\chi$ model: top-left, top-right, bottom-left, bottom-right panels: analogous to Figs.~\ref{fig:br-psi}, \ref{fig:sigma-psi}, \ref{fig:bounds-psi}, \ref{fig:sigmaLL-psi}, respectively.
}
\label{Fig:chi}
\end{figure}

\begin{figure}[t]
\begin{center}
\includegraphics[scale=0.42]{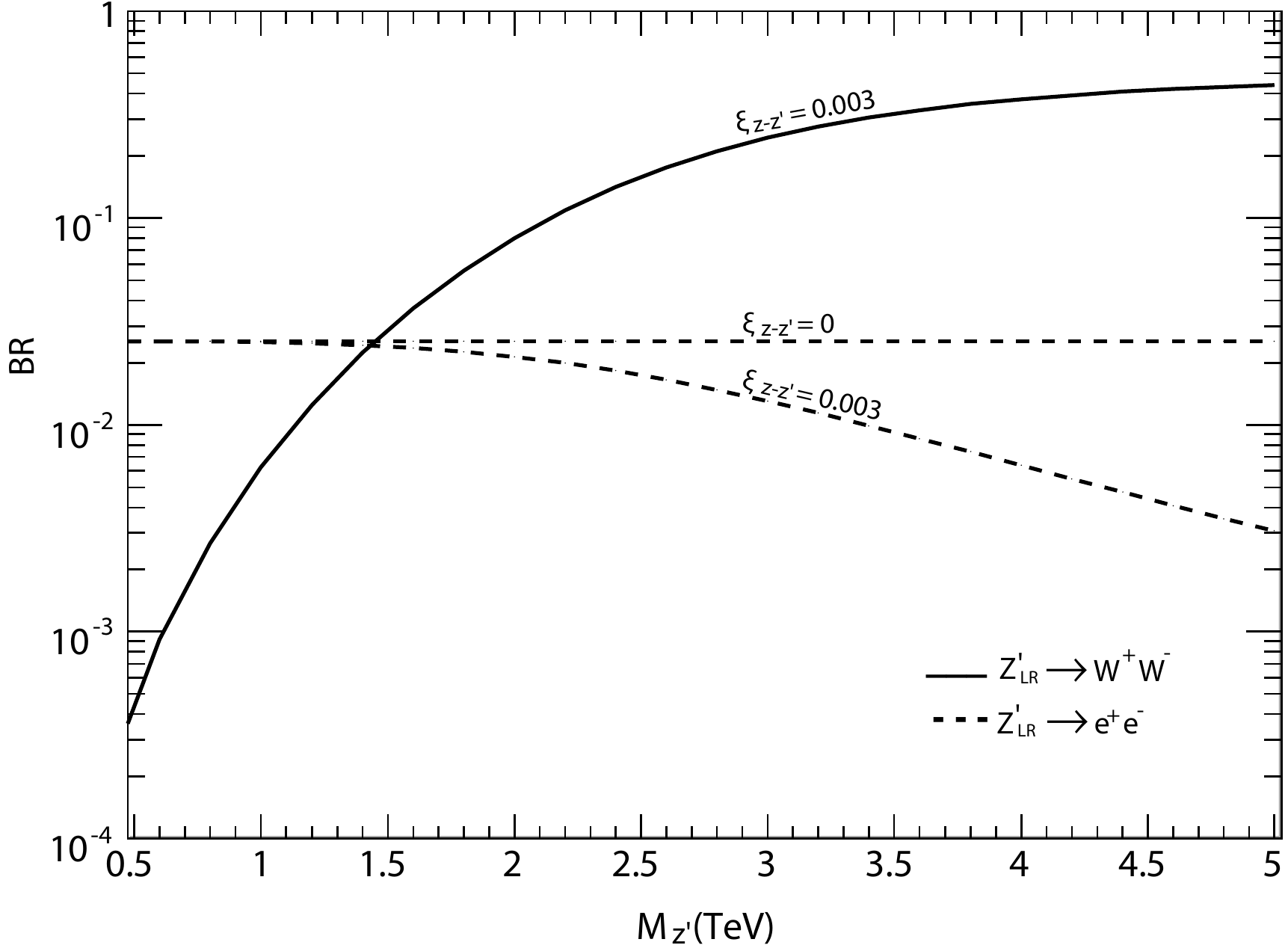}
\includegraphics[scale=0.42]{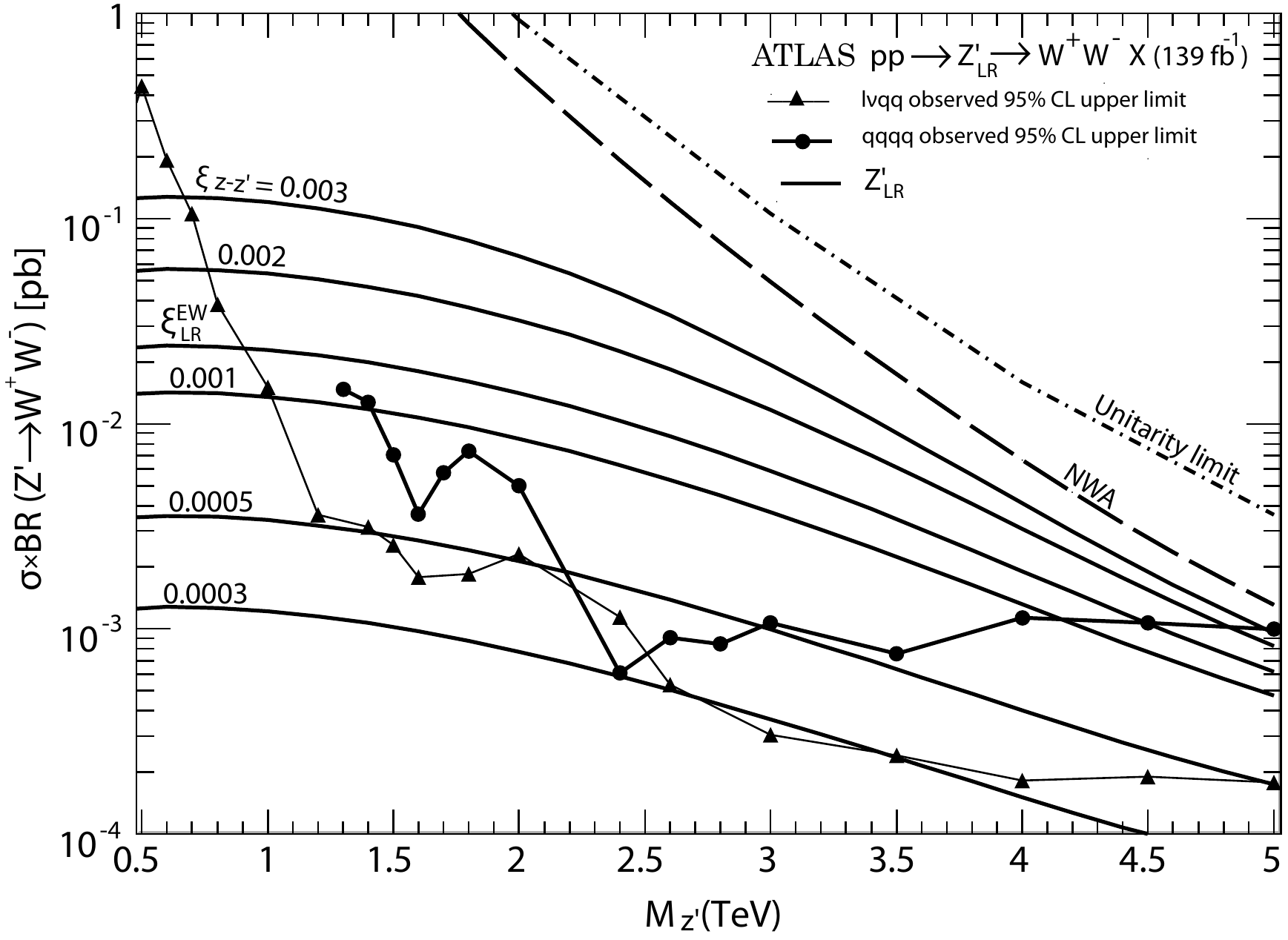}
\includegraphics[scale=0.42]{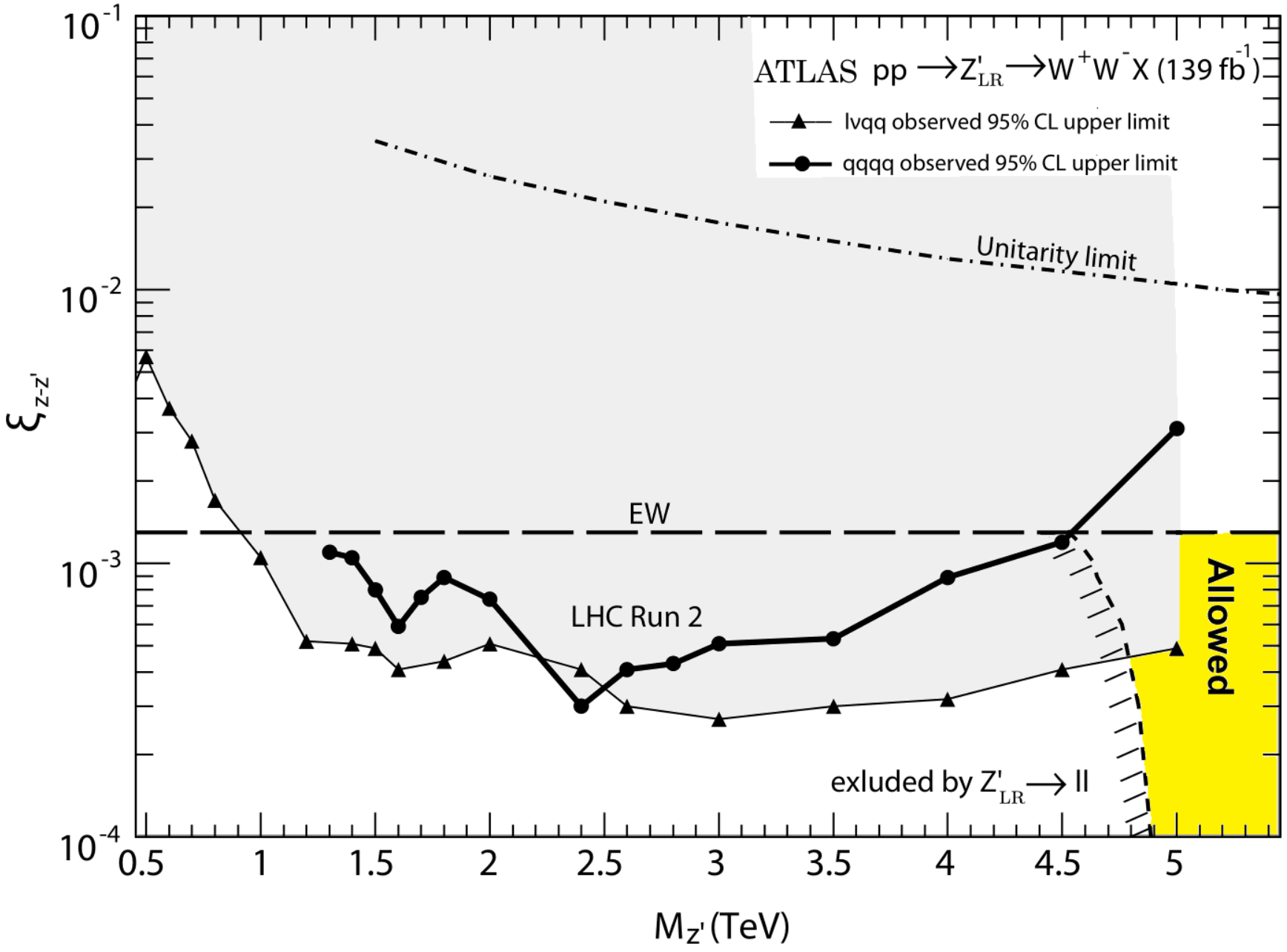}
\includegraphics[scale=0.42]{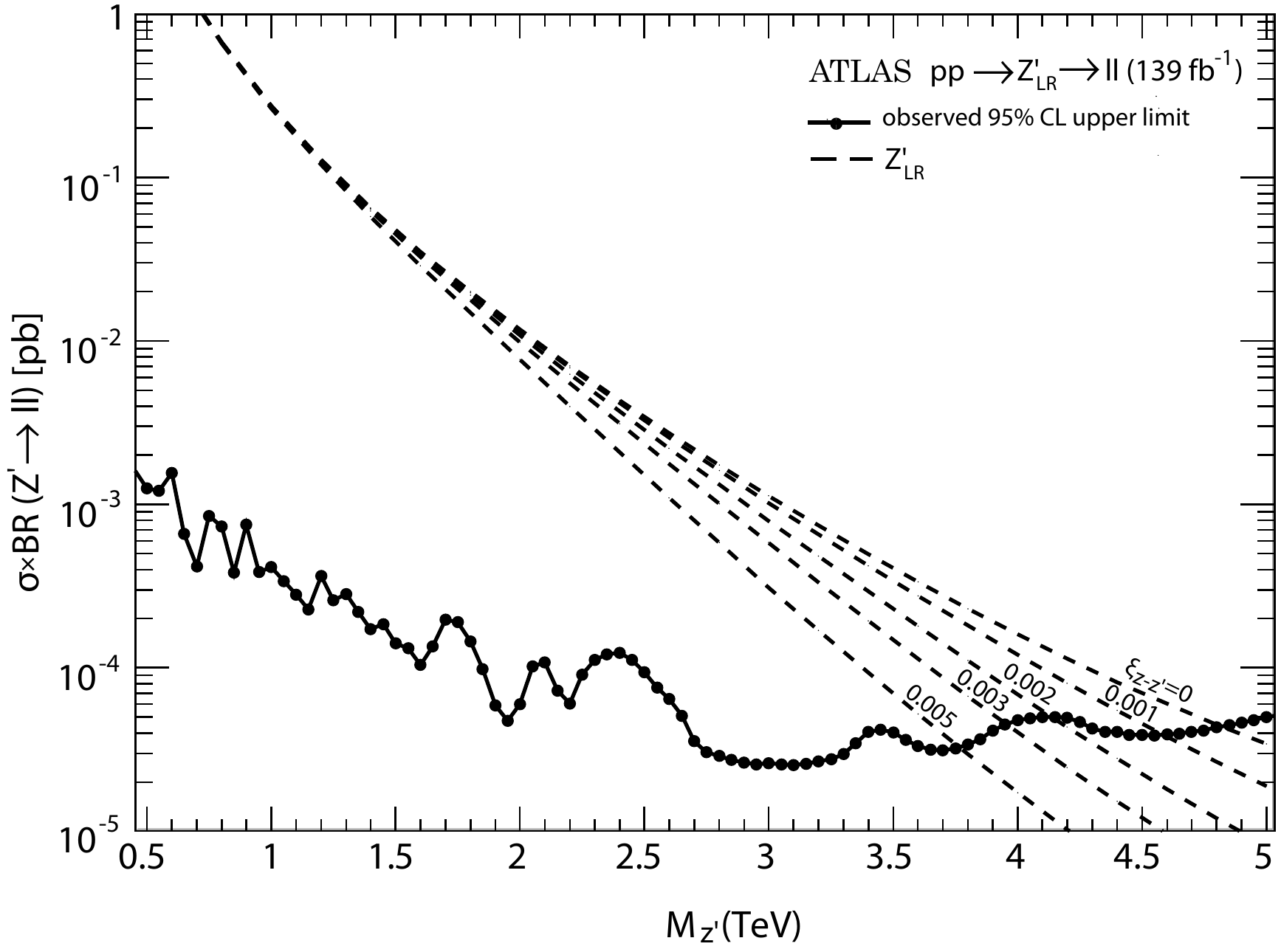}
\end{center}
\caption{The $Z'_{\rm LR}$ model: top-left, top-right, bottom-left, bottom-right panels: analogous to Figs.~\ref{fig:br-psi}, \ref{fig:sigma-psi}, \ref{fig:bounds-psi}, \ref{fig:sigmaLL-psi}, respectively.
}
\label{Fig:LR}
\end{figure}

\begin{figure}[t]
\begin{center}
\includegraphics[scale=0.42]{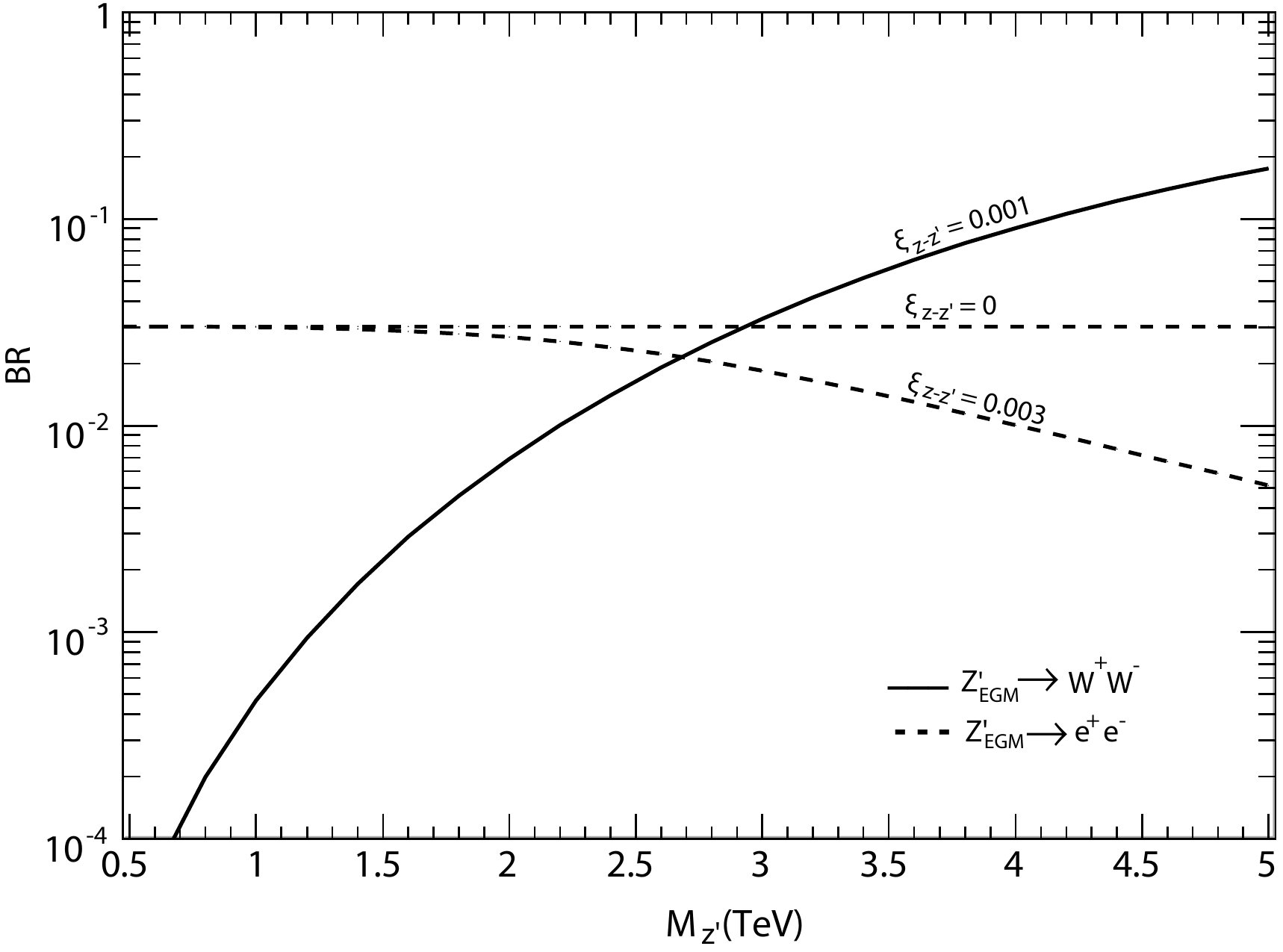}
\includegraphics[scale=0.42]{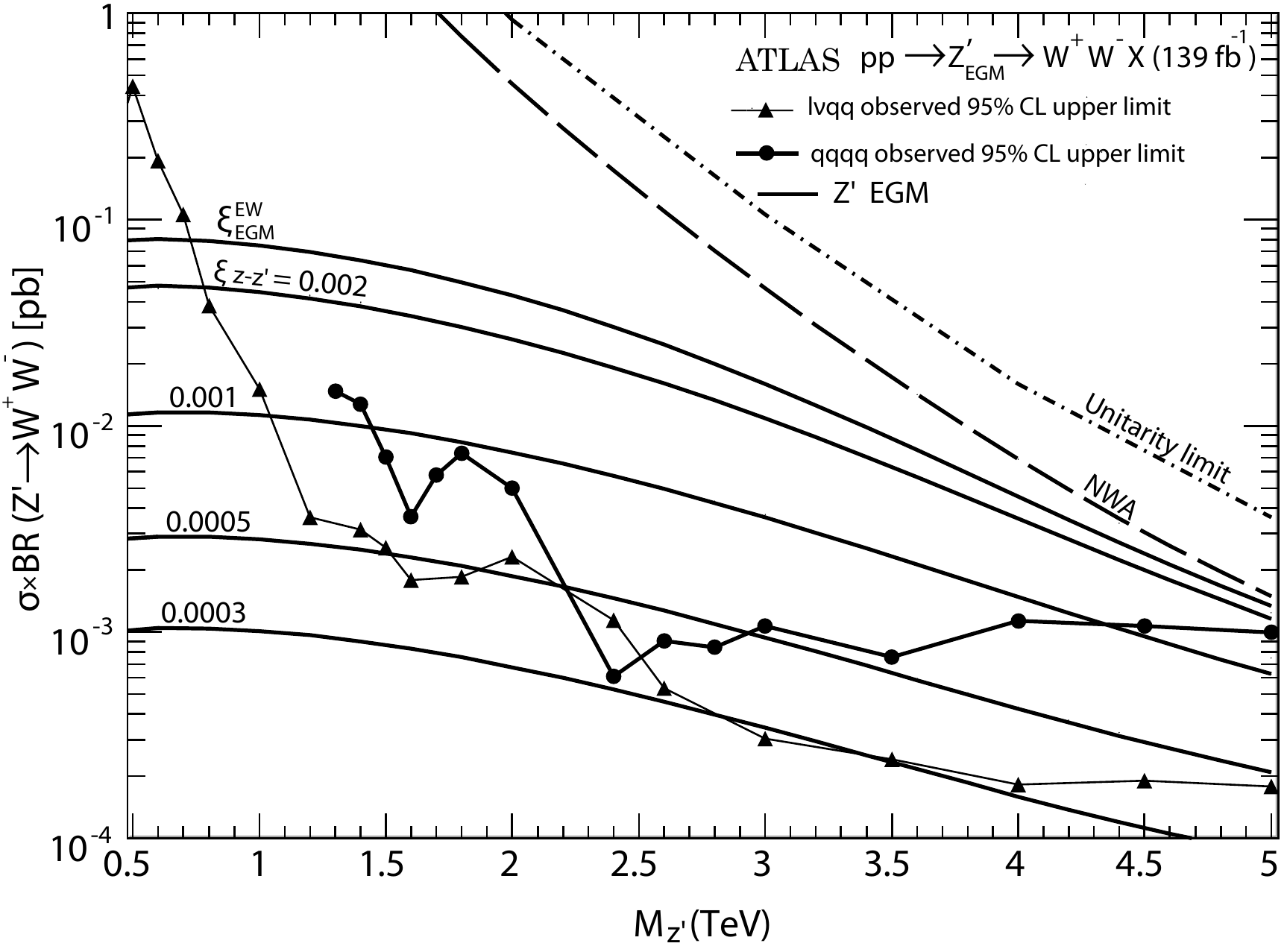}
\includegraphics[scale=0.42]{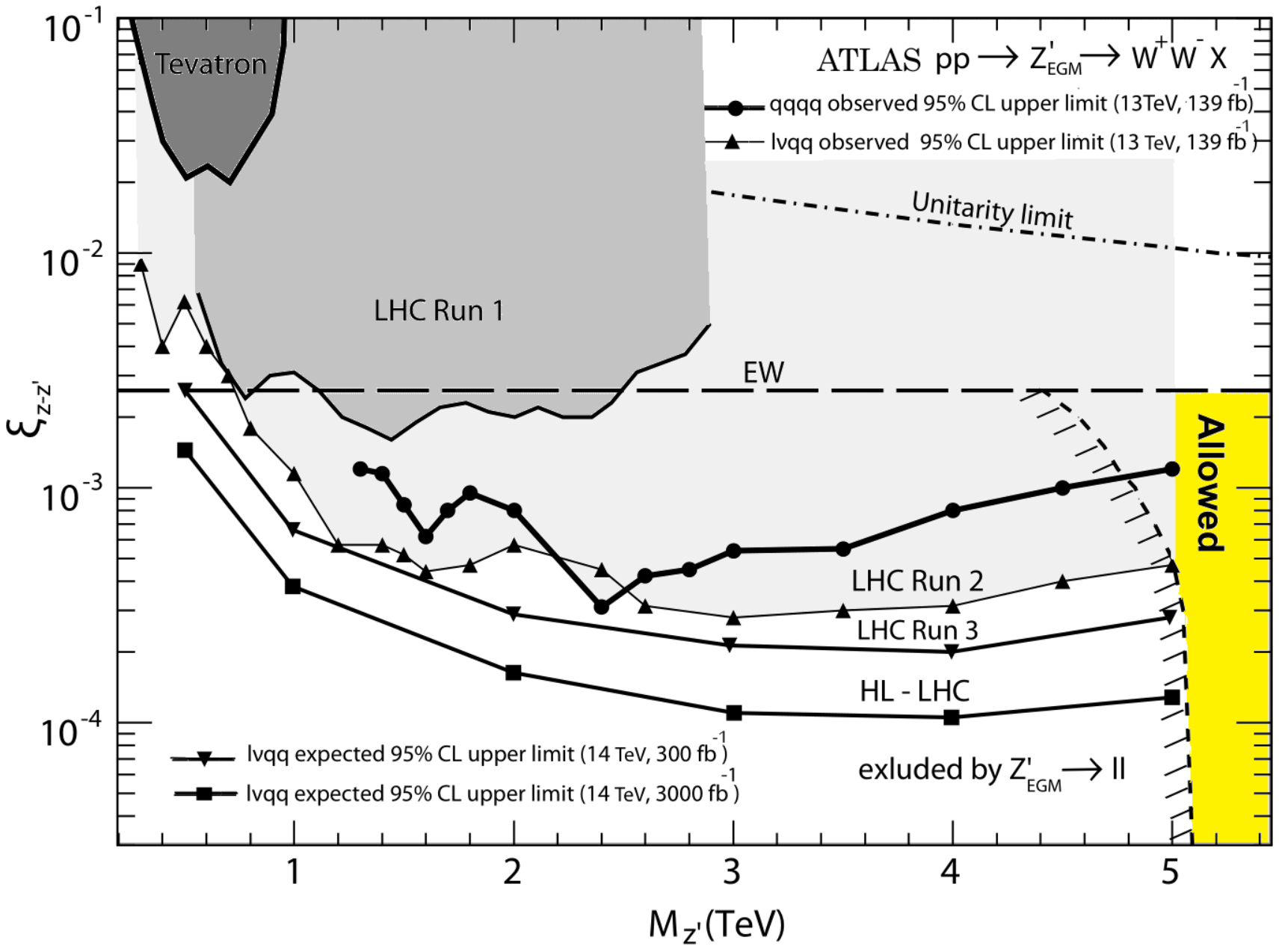}
\includegraphics[scale=0.42]{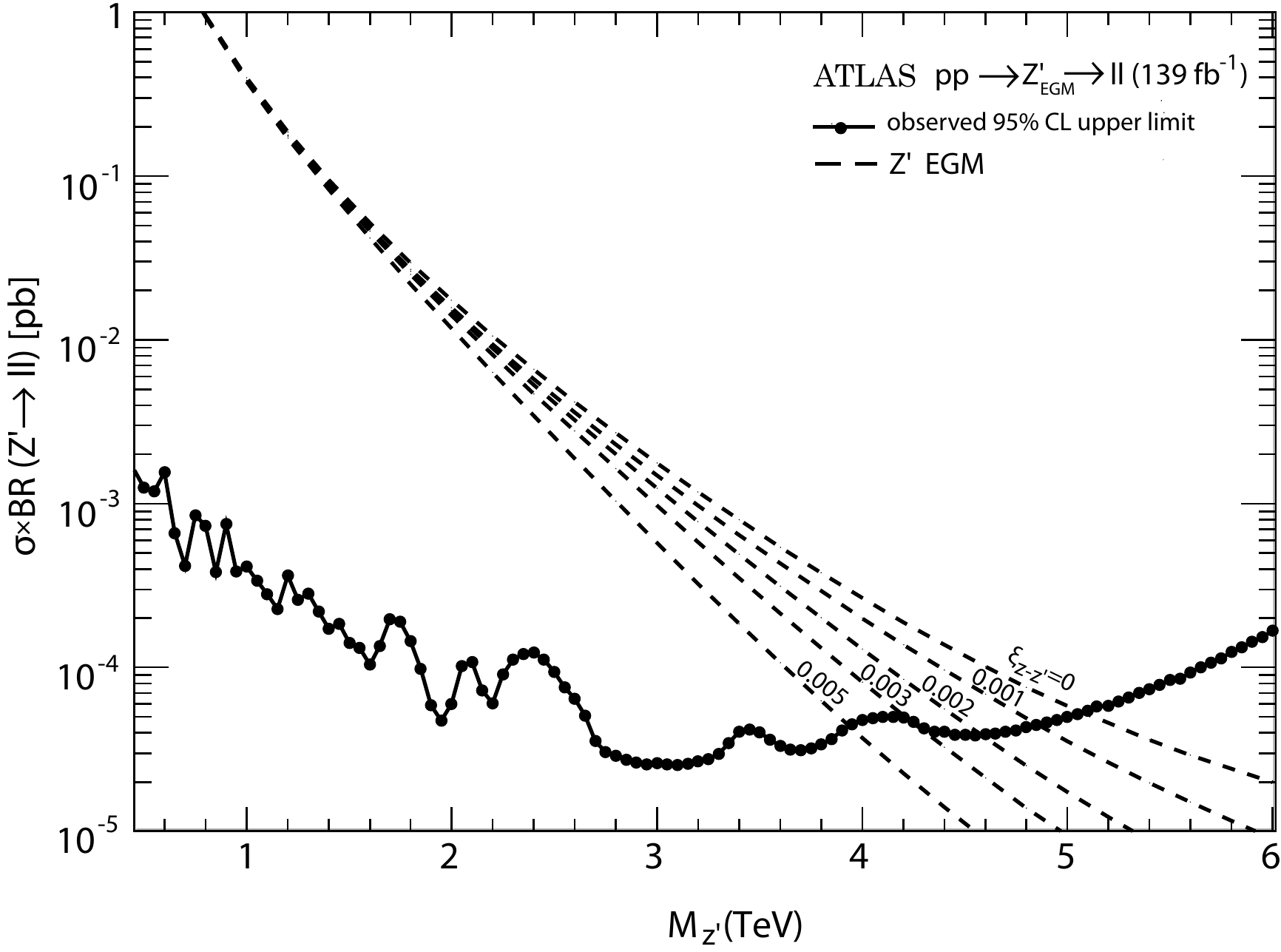}
\end{center}
\caption{The $Z'_{\rm EGM}$ model: top-left, top-right, bottom-left, bottom-right panels: analogous to Figs.~\ref{fig:br-psi}, \ref{fig:sigma-psi}, \ref{fig:bounds-psi}, \ref{fig:sigmaLL-psi}, respectively.
Bottom-left panel: Also shown are exclusion regions
obtained after incorporating direct search constraints from the CDF and D0
collaborations which are referred to as Tevatron (the dark shaded
area) in $p\bar{p}\to W^+W^-X$  as well as those derived   from
the LHC  measurement of $p{p}\to WWX$ in Run~1 (the gray area)
\cite{ Pankov:2019yzr}.
For comparison, we also show the expected exclusion from Run~3 (300~$\text{fb}^{-1}$) and the HL-LHC option (3000~$\text{fb}^{-1}$), see text.
}
\label{Fig:EGM}
\end{figure}

Comparison of $\sigma(pp\to Z'X)\times \text{BR}(Z'\to \ell\ell)$  vs $\sigma_{95\%} \times {\rm BR}(Z'\to \ell\ell)$ displayed in Fig.~\ref{sigmaLL-psi} permits us to read off an allowed mixing
for a given mass value, higher masses are allowed for smaller mixing, for the reason stated above.
This analysis of $Z$-$Z'$ mixing, illustrated here for the $\psi$ model, can also be
performed for the other benchmark models. The results of the numerical analysis for these tested models are  presented in Figs.~\ref{Fig:eta}--\ref{Fig:EGM}. Mass limits are calculated as the intersection between the observed limits with the model prediction. Table~\ref{Tab:disc} lists the mass limits for two representative cases, namely for vanishing mixing ($\xi_{Z\text{-}Z'}=0$) and for the mixing $\xi_{Z\text{-}Z'}^{\rm EW}$ derived from the electroweak precision data \cite{Erler:2009jh}. The former are consistent with those derived in Refs.~\cite{Aad:2019fac, CMS:2019tbu}  whereas the mass limits at $\xi_{Z\text{-}Z'}^{\rm EW}$ are weaker by $\sim 10-30\%$.

\begin{table}[htb]
\caption{Upper limits  on  mixing parameters $\xi_{Z\text{-}Z^\prime}$ and
$\xi_{W\text{-}W^\prime}$  at 95\% C.L.  in different models, processes
and experiments (past - Tevatron, present - EW and LHC, future - ILC).
We also compare with the expected ILC reach.
}
\begin{center}
\begin{tabular}{|c|c|c|c|c|c|c|c|}
\hline
collider, process & $\xi_{Z\text{-}Z'}^{\psi}$ & $\xi_{Z\text{-}Z'}^{\eta}$ &
$\xi_{Z\text{-}Z'}^{\chi}$ & $\xi_{Z\text{-}Z'}^{\rm LR}$ &
$\xi_{Z\text{-}Z'}^{\rm
EGM}$ &$\xi_{W\text{-}W'}^{\rm EGM}$ & @$ M_V'$ (TeV) \\ \hline
Tevatron, $ p\bar{p} \to Z'/W'   \to WW/WZ\,(\to \ell\nu\,qq)$
& ... & ...  &
... & ... & $2\cdot 10^{-2}$ & $2\cdot 10^{-2}$ & 0.4--0.9 \\
\cite{Aaltonen:2010ws}  &&&&&&&
\\ \hline
electroweak (EW) data \cite{Erler:2009jh,Zyla:2020zbs}
& $1.8 \cdot 10^{-3}$ & $4.7\cdot 10^{-3}$ & $1.6 \cdot 10^{-3}$ &
$1.3\cdot 10^{-3}$  &
$ 2.6\cdot 10^{-3}$ & $\sim  10^{-2}$ & ...
\\ \hline
LHC@13~TeV, 139 fb$^{-1}$: Run~2 (this work) &&&&&&& \\
$p{p} \to  Z'/W'\to WW/WZ \, (\to qqqq)$
&  $2{.}9 \cdot 10^{-4}$ & $2{.}7\cdot 10^{-4}$  & $3{.}8 \cdot 10^{-4}$
& $3{.}0\cdot 10^{-4}$  & $3{.}1\cdot 10^{-4}$ & $4.3 \cdot 10^{-4}$ &
1.3--5.0 \\
$p{p} \to  Z'\to WW\,
 (\to \ell\nu\, qq)$
& $2{.}5\cdot 10^{-4}$ & $2{.}4 \cdot 10^{-4}$ & $3{.}3\cdot 10^{-4}$ &
$2{.}7\cdot 10^{-4}$ & $ 2{.}8 \cdot 10^{-4}$ & ... & 0.5--5.0  \\
$p{p} \to  W'\to WZ\,
 (\to \ell\nu\,/  \ell\ell/ \nu\nu\,\, qq)$
&...&...&...&...&...& $2.9\cdot 10^{-4}$ & 0.5--5.0
\\ \hline
ILC@0{.}5~TeV, 0.5 ab$^{-1}$, $e^+e^- \to W^+W^-$
\cite{Andreev:2012cj}
& $2.3\cdot10^{-3}$ & $1.6\cdot 10^{-3}$ & $1.5\cdot 10^{-3}$ & $1.4\cdot
10^{-3}$ &  $1.2\cdot10^{-3}$ & ... & $\geq$ 3
\\
ILC@1.0~TeV, 1.0 ab$^{-1}$, $e^+e^- \to W^+W^-$
\cite{Andreev:2012cj}
& $0.6\cdot10^{-3}$ & $0.5\cdot 10^{-3}$ & $0.4\cdot10^{-3}$ &
$0.4\cdot10^{-3}$ &  $0.3\cdot10^{-3}$ & ... & $\geq$ 3
\\
\hline
\end{tabular}
\end{center}
\label{Tab:summary}
\end{table}

As described above, both the diboson mode and the dilepton process
yield limits on the ($M_{Z'}$, $\xi_{Z\text{-}Z'}$) parameter space. These are
rather complementary, as shown in Fig.~\ref{bounds-psi}, where we
collect these  limits for the $\psi$ model. The
limits arising from the diboson channel are basically excluding
large values of $\xi_{Z\text{-}Z'}$, strongest at intermediate masses
$M_{Z'}\sim 2-4~\text{TeV}$. The limits arising from the dilepton
channel, on the other hand, basically exclude masses $M_{Z'}\lsim
 4.5~\text{TeV}$, with only a weak dependence on $\xi_{Z\text{-}Z'}$.
For reference, we plot also a curve labelled ``Unitarity limit'' that corresponds to the
unitarity bound \cite{Alves:2009aa,Bobovnikov:2018fwt}. In \cite{Alves:2009aa},  it was shown that the saturation of unitarity in the elastic scattering $W^+W^-\to W^+W^-$ leads to
the constraint $(g_{Z'WW})_\text{max}=g_{ZWW}\cdot (M_Z/\sqrt{3}M_{Z'})$
that was adopted here.

For comparison, we show in Fig.~\ref{Fig:EGM} (for the EGM model) also the exclusion reach expected at the end of Run~3 (300~$\text{fb}^{-1}$) at the LHC, as well as at the HL-LHC (3000~$\text{fb}^{-1}$) \cite{CidVidal:2018eel}  which illustrates the corresponding extension of the excluded limits on $\xi_{Z-Z'}$ down to $2.0\cdot 10^{-4}$
and $1.1\cdot 10^{-4}$, respectively, whithin the $Z'$ mass range under study.
Furthermore, the expected lower limit on the $Z'$ mass can be set from the dilepton production at higher luminosity.
The current $Z_{\rm EGM}^\prime$ mass limit of 5.1 TeV at $\xi_{Z-Z'}=0$ obtained using 139 fb$^{-1}$ of data will extend to 6.7~TeV \cite{ATLAS:2018tvr}.

In Table~\ref{Tab:summary}, we collect our limits on the $Z'$ parameters
for the benchmark models. Also shown in Table~\ref{Tab:summary} are the
limits on the $Z-Z'$ ($W-W'$) mixing parameter $\xi_{Z-Z'}$ ($\xi_{W-W'}$)
from studies of diboson $WW$ ($WZ$) pair production at the Tevatron.
Table~\ref{Tab:summary} shows that the limits on $\xi_{Z-Z'}$ from the
EW precision data are generally competitive with the future
collider, ILC@0.5 TeV, but in many cases, they are stronger than
those from the Tevatron.

The diboson production at the LHC@13~TeV allows to place stringent
constraints on the $Z$-$Z^\prime$ mixing angle and $Z^\prime$ mass, $M_{Z^\prime}$. We
imposed limits on the mass and the $Z$-$Z^\prime$ mixing angle of the
$Z'$ bosons by using data comprised of $pp$ collisions at
$\sqrt{s}=13$ TeV and recorded by the ATLAS detectors at the CERN
LHC, with integrated luminosities of $\sim$~139 fb$^{-1}$ from Run~2 data
taking.

Also, we show that the derived
constraints on the $Z$-$Z^\prime$ mixing angle for the benchmark models
are of the order of  a few$\,\times 10^{-4}$   and they are greatly
improved with respect to those derived from the global analysis of
electroweak data, EW. In addition, we demonstrated in Fig.~\ref{Fig:EGM}
that further improvement on the
constraining of this mixing can be achieved from the analysis of
data to be collected at higher luminosity expected in the Run~3 and HL-LHC
options. We also 
show that only the future $e^+e^-$ linear collider ILC with
polarized beams and with very high energy and luminosity,
$\sqrt{s}=1$ TeV and $\Lumint=1\, {\rm ab}^{-1}$, may have a
chance to compete with the current LHC sensitivity to the mixing angle in
Run~2 but will not reach the levels of the Run~3 and HL-LHC options.

\section{$W'$ production and decay in $pp$ collision}
\label{sect:productionWp}
In contrast to the rich spectrum of $Z'$ models considered above, with different vector and axial-vector couplings, for $W'$ we consider only $V-A$ couplings to fermions.
\subsection{$W'$ resonant production cross section}

We consider the simplest EGM model which predicts charged heavy gauge bosons.
The analysis of $W\text{-}W'$ mixing in diboson and dilepton pair production  which will be performed below is quite analogous to that carried out in previous sections for $Z\text{-}Z'$ mixing.
At lowest order in the EGM, $W'$ production and decay into $WZ$ in
proton-proton collisions occurs through quark-antiquark
annihilation in the $s$-channel.
Using the NWA, one can factorize the
process (\ref{procWZ}) into the $W'$ production and the $W'$
decay,
\begin{equation}
\sigma(pp\to W' X\to WZX)  = \sigma(pp\to W'X) \times \text{BR}(W' \to
WZ)\;.
\label{TotCr}
\end{equation}
Here, $\sigma(pp\to W' X)$ is the  total (theoretical) $W'$
production cross section and
$\text{BR}(W' \to WZ)=\Gamma_{W'}^{WZ}/\Gamma_{W'}$ with
$\Gamma_{W'}$ the total width of the $W'$.

\subsection{The $W'$ width}
\label{sect:width}

In the EGM the $W'$ bosons can decay into SM fermions, gauge
bosons ($WZ$), or $WH$. In the calculation
of the total width $\Gamma_{W'}$ we consider the following
channels: $W'\to f{\bar{f}}^\prime$, $WZ$, and $WH$, where
$f$ is a SM fermion ($f=\ell,\nu,q$) \footnote{Here, in contrast to the $Z'$ case, the $\ell$ includes $\tau$ leptons.}.
Only left-handed neutrinos are considered,
possible right-handed neutrinos are assumed to be kinematically
unavailable as final states. Also, like for the $Z'$ case, we shall
ignore the couplings to other beyond-SM particles such as
SUSY partners and exotic fermions.
 As a result, the total decay
width of the $W^\prime$ boson is taken to be
\begin{equation}\label{gamma}
\Gamma_{W'} = \sum_f \Gamma_{W'}^{f\bar{f}'} + \Gamma_{W'}^{WZ} +
\Gamma_{W'}^{WH}.
\end{equation}
Like for the $Z'$ case, the presence of the last two decay channels,  which are often
neglected at low and moderate values of $M_{W'}$, is due to
$W$-$W'$ mixing which is constrained to be tiny.
In particular, for the
range of $M_{W'}$ values below $\sim 1.0-1.5$ TeV,  the dependence
of $\Gamma_{W'}$ on the values of $\xi_{W\text{-}W^\prime}$ (within its allowed range)
induced by $\Gamma_{W'}^{WZ}$ and $\Gamma_{W'}^{WH}$ is
unimportant because $\sum_f \Gamma_{W'}^{f\bar {f'}}$ dominates
over the diboson partial widths. Therefore, in this mass range, one
can approximate the total width as $\Gamma_{W'} \approx \sum_f
\Gamma_{W'}^{f\bar {f'}}=3.5\%\times M_{W'}$ \cite{Serenkova:2019zav},
 where the sum runs over SM fermions only.

For heavier $W'$ bosons, the diboson decay channels, $WZ$ and
$WH$, start to play an important role,
and we are no longer able to ignore them \cite{Serenkova:2019zav,Pankov:2019yzr}.
To be specific, in analogy with the $Z'$ case, we assume that
both partial widths are comparable, $\Gamma_{W'}^{WH}\simeq
\Gamma_{W'}^{WZ}$ for heavy $M_{W'}$, as required by the
Equivalence theorem \cite{Chanowitz:1985hj}.

\begin{figure}[t]
\begin{center}
\includegraphics[scale=0.49]{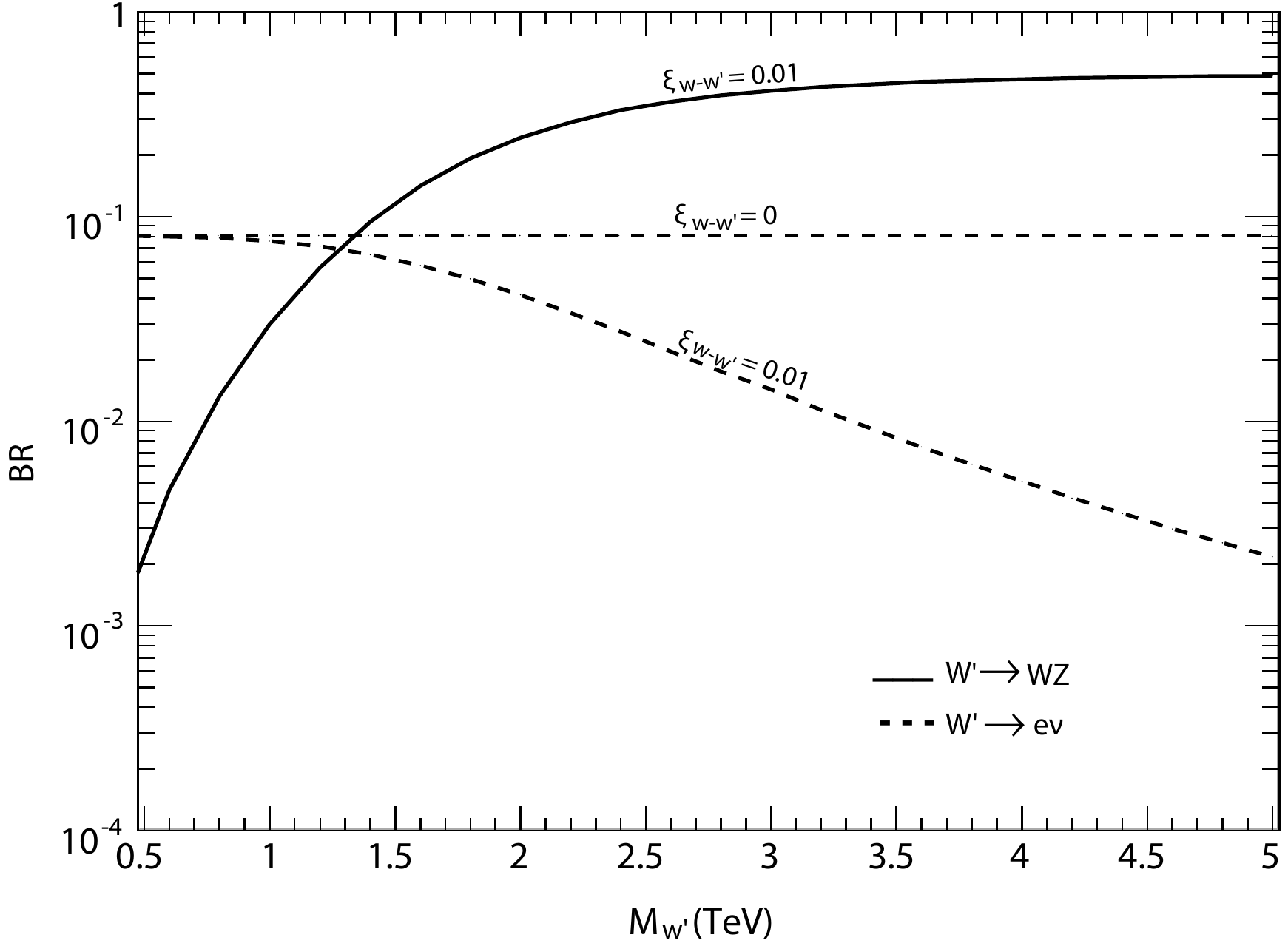} \ \ \
\includegraphics[scale=0.49]{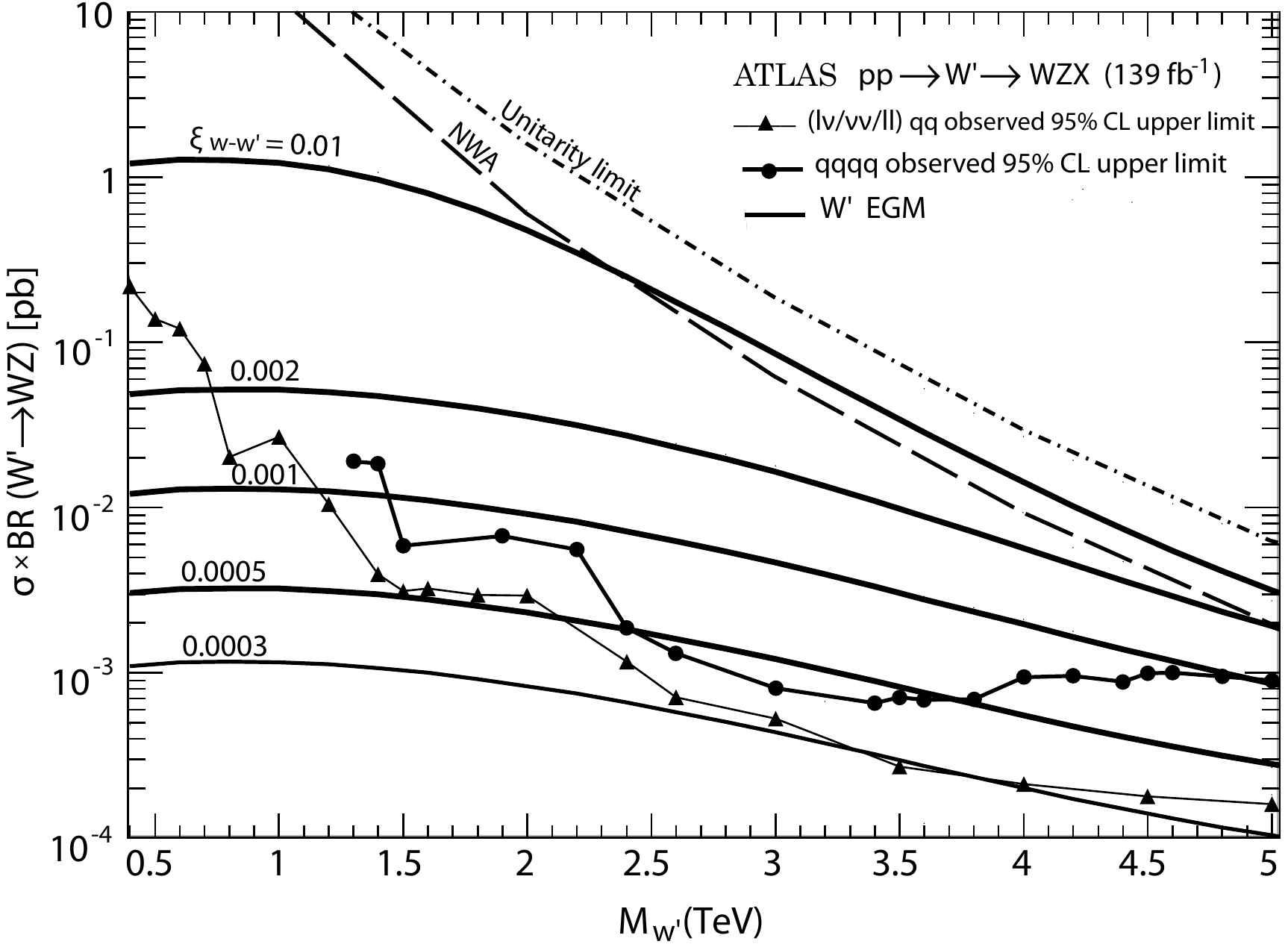}
\end{center}
\caption{
Left panel: Branching ratio $\text{BR}(W'\to WZ)$ (solid) vs
$M_{W'}$ in the EGM for $W$-$W'$ mixing factor
$\xi_{W\text{-}W^\prime}=10^{-2}$.
 Dashed: $\text{BR}(W'\to e\nu)$ for
$\xi_{W\text{-}W'}=0$ ($W'_{\rm SSM}$) and $\xi_{W\text{-}W'}=0.01$.
Right panel:
$95\%$ C.L. upper limits on $\sigma_{95\%}\times \text{BR}(W'\to WZ)$, 
showing ATLAS data on the fully
hadronic and semileptonic final states for  $139~\text{fb}^{-1}$ \cite{Aad:2019fbh,Aad:2020ddw}.
The theoretical production cross sections $\sigma(pp\to W'X)\times \text{BR}(W'\to
WZ)$ for the ${\rm EGM}$ are calculated from PYTHIA with a $W'$
mass-dependent $K$-factor,
given by solid curves, for mixing factor $\xi_{W\text{-}W^\prime}$
ranging from $10^{-2}$ and down to $3\cdot 10^{-4}$.  The NWA and
unitarity constraints are also shown \cite{Alves:2009aa, Serenkova:2019zav}.
}
\label{Fig:Wprime1}
\end{figure}

The expression for the partial width of the $W'\to WZ$ decay
channel in the EGM can be written as \cite{Altarelli:1989ff,Serenkova:2019zav}:
\begin{eqnarray}
\label{GammaWZ}
 \Gamma_{W'}^{WZ}&=&\frac{\alpha_{\rm
em}}{48}\cot^2\theta_W\, M_{W'}
\frac{M_{W'}^4}{M_W^2M_Z^2}\left[\left(1-\frac{M_Z^2-M_W^2}{M_{W'}^2}\right)^2
-4\,\frac{M_W^2}{M_{W'}^2}\right]^{3/2} \\ \nonumber
&& \times\left[
1+10 \left(\frac{M_W^2+M_Z^2}{M_{W'}^2}\right) +
\frac{M_W^4+M_Z^4+10M_W^2M_Z^2}{M_{W'}^4}\right]\cdot\xi_{W\text{-}W'}^2.
\end{eqnarray}

For a fixed mixing factor $\xi_{W\text{-}W^\prime}$  and at large $M_{W'}$, 
the total width
increases rapidly with the $W'$ mass because of the quintic
dependence  of the $WZ$ mode on the $W'$ mass
$\Gamma_{W'}^{WZ}\propto M_{W'}\left[{M_{W'}^4}/({M_W^2M_Z^2})\right]$,
 corresponding to the production of longitudinally polarized $W$ and $Z$ in the channel $W'\to
W_LZ_L$ \cite{Altarelli:1989ff,Serenkova:2019zav}.
In this case, the $WZ$ mode (as well as $WH$) becomes dominant and
$\text{BR}(W' \to WZ)\to 0.5$, while the fermionic decay channels,
$\sum_f \Gamma_{W'}^{f\bar {f'}}\propto M_{W'}$, are increasingly
suppressed, as illustrated in Fig.~\ref{Fig:Wprime1} (left panel).

\subsection{Hadron production and diboson decay of $W'$}

Our analysis employs the  recent searches for
diboson processes in semileptonic final states provided by ATLAS \cite{Aad:2020ddw}
with the full Run~2 data set with time-integrated luminosity of 139 fb$^{-1}$ as well as, for the sake of comparison, in the fully hadronic ($qqqq$) final states \cite{Aad:2019fbh}.

In Fig.~\ref{Fig:Wprime1} (right panel), we show the observed
$95\%$ C.L. upper limits on the production cross section times the branching
fraction, $\sigma_{95\%}\times \text{BR}(W'\to WZ)$, as a function
of the $W'$ mass.

Then, for $W'$ we compute the LHC theoretical production cross
section multiplied by the branching ratio into  $WZ$ bosons,
$\sigma (pp\to W'X)\times {\rm BR}(W'\to WZ)$, as a function of the
two parameters ($M_{W'}$, $\xi_{W\text{-}W^\prime}$) \cite{Serenkova:2019zav}, and compare it with the limits established by the ATLAS experiment, $\sigma_{95\%} \times {\rm BR}(W'\to WZ)$.
The simulation of signals for the EGM $W'$ is based on a suitably adapted version of
the leading order PYHTHIA 8.2 event generator \cite{Sjostrand:2014zea}.
A mass-dependent $K$ factor is adopted to rescale the LO
PYTHIA prediction to the the NNLO one, using the 
ZWPROD \cite{Hamberg:1990np} software. The result is presented as solid curves
in the right panel for a mixing factor $\xi_{W\text{-}W^\prime}$ ranging from
$10^{-2}$ and down to $3\cdot 10^{-4}$. The factorization and renormalization scales are both set to the $W'$ mass.

The area below the
long-dashed curve labelled ``NWA'' corresponds to the region where
the $W'$ resonance width is predicted to be less than 5\% of its
mass, corresponding to the best detector resolution of
the searches, where the narrow-width assumption is satisfied.
We also show a curve labelled  ``Unitarity limit'' that corresponds to the unitarity bound (see, e.g. \cite{Alves:2009aa} and references therein). It was shown that the saturation of
unitarity in the elastic scattering $W^\pm Z\to W^\pm Z$ leads to
the constraint  $(g_{W'WZ})_\text{max}=g_{WWZ}\cdot M_Z^2/(\sqrt{3}\,M_{W'}\,M_W)$
that was adopted in plotting this bound. The constraint
was obtained under the assumption  that the couplings of the $W^\prime$ to
quarks and to gauge bosons have the same Lorentz structure
as those of the SM but with rescaled strength.

\begin{figure}[t]
\begin{center}
\includegraphics[scale=0.49]{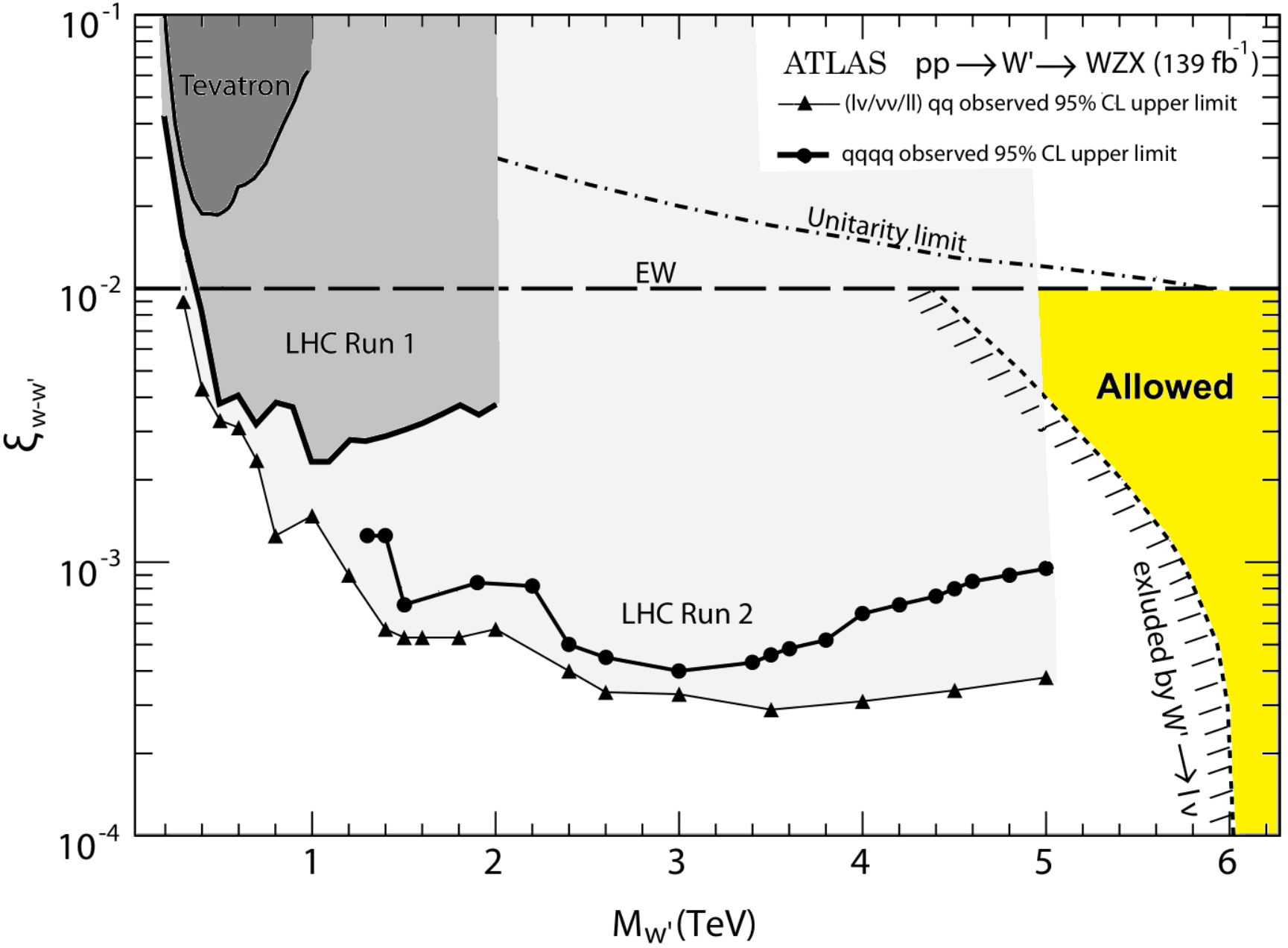}
\includegraphics[scale=0.49]{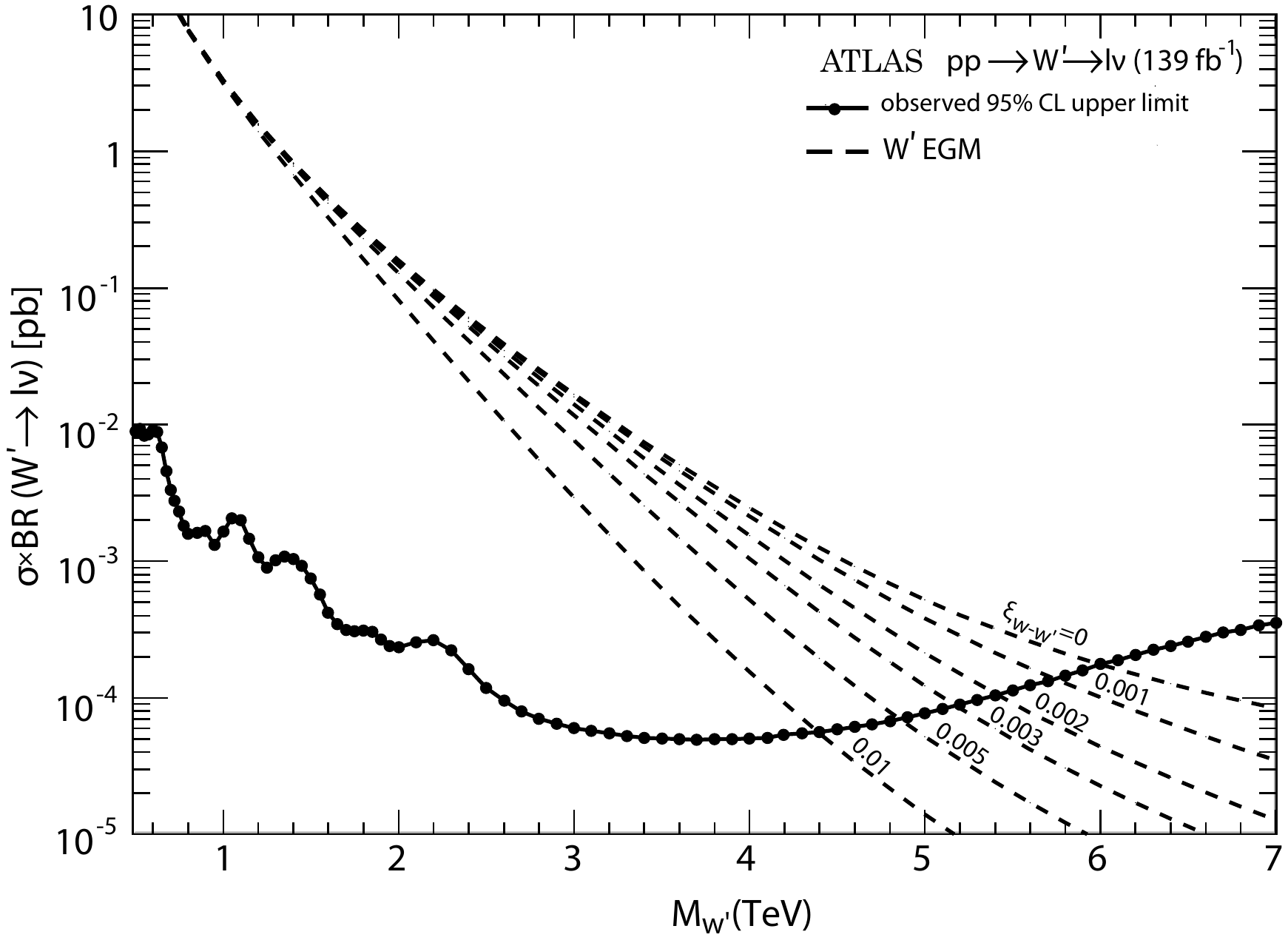}
\end{center}
\caption{
Left panel: 95\% C.L. exclusion regions  in the two-dimensional
($M_{W'}$, $\xi_{W\text{-}W^\prime}$) plane obtained from the
precision electroweak data (horizontal dashed straight line
labeled ``EW''), direct search constraints from the Tevatron in
$p\bar{p}\to WZX$ (dark shaded area) as well as from the LHC
searches for $p{p}\to WZX$ at 7 TeV and 8 TeV (Run~1) (gray
area)  and at 13~TeV from diboson $W'\to WZ$ production in hadronic and semileptonic  final states
using the  full Run~2 ATLAS data set. The region above each curve for the $WZ$ channel is excluded.
The steep curve labelled ``excluded by
$W'\to \ell\nu$'' shows the exclusion based on the dilepton channel $pp\to \ell\nu X$.
The overall allowed region for  the  EGM $W'$ boson is shown as the yellow area.  
Right panel:
Solid: observed $95\%$ C.L. upper bound on the $W'$ production cross section times
branching ratio to two leptons, $\sigma_{95\%}\times
\text{BR}(W'\to \ell\nu)$, obtained in the combined channels (electron and muon) at the LHC with integrated luminosity $\Lumint$=139\, fb$^{-1}$ by the ATLAS collaboration
\cite{Aad:2019wvl}. Dashed lines: theoretical production cross
section $\sigma(pp\to W') \times {\rm BR}(W'\to \ell\nu)$
 for the EGM $W'$ boson, calculated from PYHTHIA~8.2 with an NNLO $K$ factor.
These curves, in descending order, correspond to 
values of the $W$-$W'$ mixing factor $\xi$ from 0 to 0.01.
}
\label{Fig:Wprime2}
\end{figure}

The theoretical curves for the cross sections $\sigma(pp\to W'X)
\times {\rm BR}(W'\to WZ)$, in descending order, correspond to
values of the $W$-$W'$ mixing factor $\xi_{W\text{-}W^\prime}$ from 0.01 to 0.0003.
The intersection points of the measured upper
limits on the production cross section with these theoretical
cross sections for various values of $\xi_{W\text{-}W^\prime}$ give the corresponding lower
bounds on ($M_{W'}$, $\xi_{W\text{-}W^\prime}$), displayed  in
Fig.~\ref{Fig:Wprime2}, left panel.

 Comparison of sensitivities of the process (\ref{procWZ}) to $W'$ with different decay channels, e.g., $VV\to\ell\nu qq$ and $qqqq$, can be performed by the matching of $95\%$ C.L.\ upper limits on the production cross section times the branching fraction, $\sigma_{95\%}\times \text{BR}(W'\to WZ)$, which includes the SM branching fractions of the electroweak bosons to the final states in the analysis channel, effects from detector acceptance, as well as reconstruction and selection efficiencies. ATLAS bounds were included acccording to the Durham HEPdata repository
 \cite{hepdata1740685,hepdata1793572}.

From a comparison of the upper limits on the production cross section times the branching
fraction for semileptonic  vs.\ fully hadronic decay channels, one can  conclude that
the sensitivity of the semileptonic channel dominates over the fully hadronic one within the whole range of the $W'$ mass, from 0.5 TeV to 5 TeV. These features are illustrated in Fig.~\ref{Fig:Wprime1} (right panel) and Fig.~\ref{Fig:Wprime2} (left panel).

For reference, we display limits on the $W'$ parameters
from the Tevatron (CDF and D0) as well as from ATLAS and CMS
obtained at 7 and 8 TeV of LHC data taking in Run~1 denoted ``LHC Run~1'' \cite{Serenkova:2019zav}.
Fig.~\ref{Fig:Wprime2} (left panel) shows that the experiments CDF and D0 at the Tevatron exclude EGM $W'$ bosons with $\xi_{W\text{-}W^\prime}\gsim 2\cdot 10^{-2}$ in the resonance mass range 0.25~TeV $<M_{W'}<$ 1~TeV at the $95\%$ C.L., whereas LHC in Run~1 improved those constraints, excluding  $W'$ boson parameters at $\xi_{W\text{-}W^\prime}\gsim 2\cdot 10^{-3}$ in the mass range 0.2~TeV  $<M_{W'}<$ 2~TeV.

As expected, the increase of the time-integrated luminosity up to
139~fb$^{-1}$ leads to dominant sensitivity of the semileptonic channel over the whole resonance mass range of 0.5~TeV $<M_{W'}<$ 5~TeV and it allows to set stronger constraints on the mixing
angle $\xi_{W\text{-}W^\prime}$, excluding $\xi_{W\text{-}W^\prime} >  2.3\cdot 10^{-4}$  as shown in Fig.~\ref{Fig:Wprime2}.
Our results extend the sensitivity beyond the corresponding CDF Tevatron  results \cite{Aaltonen:2010ws}  as well as the ATLAS and CMS sensitivity attained
at 7 and 8~TeV. Also, for the first time, we set $W'$ limits as functions
of the mass $M_{W'}$ and mixing factor $\xi_{W\text{-}W^\prime}$ from the study of the diboson production and subsequent decay into semileptonic final states at the LHC at 13~TeV
with the full ATLAS Run~2 data set.
The exclusion region obtained in this way on the
parameter space of the $W'$ naturally supersedes the corresponding exclusion area obtained for
time-integrated luminosity of 36.1~fb$^{-1}$ in the semileptonic channel as reported in \cite{Serenkova:2019zav}.
The limits on the $W'$ parameters presented in this section obtained from the diboson $WZ$ production in semileptonic final states, corresponding to  a time-integrated luminosity of 139~fb$^{-1}$, are the best to date.

\subsection{$W\text{-}W'$ mixing effects in dilepton decay of $W'\to\ell\nu$}

The above analysis was for the diboson process (\ref{procWZ}),
employing one of the most recent ATLAS searches
\cite{Aad:2020ddw,Aad:2019fbh}. Next, we turn to the dilepton production
process (\ref{proclept}), this process gives valuable
complementary information. Unlike the SSM, where there is
no $W$-$W'$ mixing, in the EGM we consider a non-zero
mixing $\xi_{W\text{-}W'}$ in the analysis of
the $W'\to \ell\nu$ process. As described in Sec.~\ref{sect:width},
this results in a modification of $\text{BR}(W'\to e\nu)$.

We compute the $W'$ production cross section at LO with PYTHIA~8.2
\cite{Sjostrand:2014zea}, $\sigma(pp\to W')$,
multiplied by the branching ratio into two leptons, $\ell\nu$
(here $\ell=e$), i.e., $\sigma(pp\to W') \times {\rm BR}(W'\to
e\nu)$, as a function of $M_{W'}$.  A mass-dependent $K$
factor is applied, based on NNLO QCD cross sections as calculated
with FEWZ~3.1 \cite{Gavin:2010az,Li:2012wna}. The $K$ factor varies
approximately from 1.3 to 1.1 for the range of $W'$ masses studied
in this analysis, namely from 0.5 to 6.0~TeV. The NNLO corrections
decrease with increasing $W'$ mass up to around 4.5~TeV
\cite{Khachatryan:2016jww}. For higher $W'$ masses,
 the $K$ factor increases again and
becomes similar to the low-mass values.

The product of the NNLO $W'$  theoretical production cross section
and branching fraction, $\sigma(pp\to W') \times {\rm BR}(W'\to
e\nu)$, for the $W'$ boson for EGM strongly depends on the $W'$
mass, and is given by dashed curves,
in descending order, corresponding to values of the mixing factor
$\xi$ from 0.0 to 0.01, as displayed in Fig.~\ref{Fig:Wprime2} (right panel). 

Comparison of $\sigma(pp\to W') \times {\rm BR}(W'\to e\nu)$ vs
$\sigma_{95\%}\times \text{BR}(W'\to \ell\nu)$ displayed in
Fig.~\ref{Fig:Wprime2} (right panel) allows us to read off an allowed mixing for a
given mass value, higher masses are allowed for smaller mixing,
for the reason stated above.  That comparison can be translated
into constraints on the two-dimensional $M_{W'}$-$\xi_{W\text{-}W'}$ parameter
plane, as  shown in Fig.~\ref{Fig:Wprime2} (left panel).

The above  results are based on  data
corresponding to an integrated luminosity of 139 fb$^{-1}$ taken
by the ATLAS  collaboration at $\sqrt{s} = 13$ TeV in Run~2  \cite{Aad:2019wvl}.
The corresponding lower limits on the $W'$ boson mass of 6 TeV (at $\xi_{W\text{-}W'}=0$) was set at 95\% C.L. from combination of the electron and muon channels. Notice that, similar to the case of $Z'$ bosons, at $\xi_{W\text{-}W'}^{\rm EW}=10^{-2}$ these limits  become weaker, reaching $\sim$ 4.4 TeV, as  illustrated in Fig.~\ref{Fig:Wprime2} (left and right panels).

\section{Concluding remarks}
\label{sect:conclusions}

Examination of the diboson, $WW$ and $WZ$, and dilepton, $\ell\ell$ and $\ell\nu$, production at the LHC with the 13~TeV data set allows to place stringent constraints on the $Z$-$Z'$ and $W$-$W'$  mixing parameters as well as on the $Z'$ and $W'$ masses for benchmark extended models, respectively. We derived such limits by using the full ATLAS Run~2 data set recorded at the CERN LHC, with integrated luminosity of 139~fb$^{-1}$. The constraints are summarized in Table~\ref{Tab:summary}. We note that in a situation when the limit is dominated by statistical errors, the $K$-factor plays a role similar to integrated luminosity. Thus, if we had adopted the same $K$-factor for the $WW$ ($WZ$) channel as for the dilepton channel, the bounds on $\xi$ would have been slightly weaker, by a factor $K^{1/4}\simeq1.17$ for the $Z'$ case \cite{Bobovnikov:2018fwt}.

By comparing the experimental limits to the theoretical predictions for the total cross section of the $Z'$ and $W'$ resonant production and their subsequent decays into $WW$ or $WZ$ pairs, we
show that the  derived constraints on the mixing parameters, $\xi_{Z\text{-}Z^\prime}$ and $\xi_{W\text{-}W^\prime}$,  are substantially improved
with respect to those obtained from the global analysis of low
energy electroweak data, as well as compared to the diboson production study
performed at the Tevatron, and to those published previously and based on the LHC Run~1 as well as at 13 TeV in Run~2 at time-integrated luminosity of $\sim 36$ fb$^{-1}$ and are the most stringent bounds to date. Further constraining of this mixing can be achieved from the analysis of data
to be collected in Run~3 as well as at the next options of hadron colliders such as HL-LHC and HE-LHC
\cite{CidVidal:2018eel,Abada:2019ono}.

\vspace*{0mm}

\section*{Acknowledgements}
It is a pleasure to thank Dr. Anurag Tripathi for discussions on $K$-factors.
This research has been partially supported by the Abdus Salam ICTP
(TRIL Programme) and by the Belarusian
Republican Foundation for Fundamental Research, F20MC-004 and F20MC-005.
The work of PO has been supported by the
Research Council of Norway.


\bibliography{ref}

\end{document}